\documentclass[12pt]{article}
\usepackage{scicite}
\usepackage{times}
\usepackage{amsmath}
\usepackage{amsfonts}
\usepackage{amssymb}
\usepackage{graphicx}
\usepackage{textcomp}

\newenvironment{sciabstract}{%
\begin{quote} \bf}
{\end{quote}}

\title{Diffuse Scattering from Correlated Electron Systems} 

\author{Raymond Osborn,${}^{1\ast}$ Damjan Pelc,${}^{2,3}$, 
Matthew J. Krogstad,${}^{4}$\\Stephan Rosenkranz,${}^{1}$
Martin Greven${}^{2}$ 
\\
\normalsize{${}^{1}$Materials Science Division, Argonne National Laboratory,
Lemont, IL 60439, USA}\\
\normalsize{${}^{2}$School of Physics and Astronomy, University of Minnesota,
Minneapolis, MN 55455, USA}\\
\normalsize{${}^{3}$Department of Physics, Faculty of Science, University of
Zagreb, Zagreb HR-10000, Croatia}\\
\normalsize{${}^{4}$Advanced Photon Source, Argonne National Laboratory, Lemont,
IL 60439, USA}\\
\\
\normalsize{$^\ast$To whom correspondence should be addressed; 
E-mail: rosborn@anl.gov.}
}

\date{}

\begin{document} 
\baselineskip24pt
\maketitle

\begin{sciabstract}
  The role of inhomegeneity in determining the properties of correlated electron
  systems is poorly understood because of the dearth of structural probes of
  disorder at the nanoscale. Advances in both neutron and x-ray scattering
  instrumentation now allow comprehensive measurements of diffuse scattering in
  single crystals over large volumes of reciprocal space, enabling structural
  correlations to be characterized over a range of length scales from 10~Å to
  200~Å or more. When combined with new analysis tools, such as
  three-dimensional difference pair-distribution functions, these advanced
  capabilities have produced novel insights into the interplay of structural
  fluctuations and electronic properties in a broad range of correlated electron
  materials. This review describes recent investigations that have demonstrated
  the importance of understanding structural inhomogeneity pertaining to
  phenomena as diverse as superconductivity, charge-density-wave modulations,
  metal-insulator transitions, and multipolar interactions.
\end{sciabstract}

\section*{Introduction}

It has long been recognized that crystalline inhomogeneity profoundly influences
the properties of correlated electron materials~\cite{Aeppli.1997ju,
Dagotto.2005ip, Keen.2015bq, Mazza.2024}. On the one hand, structural disorder
can have a deleterious effect on long-range electronic order or the formation of
spin-liquid ground states~\cite{Zhou.2017dk, Broholm.2020cc}. On the other hand,
it can be pivotal to generating emergent electronic order, for example, in the
case of substitutional doping to tune charge-carrier
densities~\cite{Tokura.2006ff, Lee.2006de, Canfield.2010} or by inducing
quantum fluctuations close to a quantum phase transition~\cite{Vojta.2003,
vanderMarel.2003cf, Shibauchi.2014bw}. However, the mechanisms that underlie the
influence of structural inhomogeneity on electronic correlations are often
poorly understood, so theoretical approaches tend to fall into two camps.
Crystalline disorder is either treated as randomly substituted point defects
within a rigid-band model or, conversely, as phase-separated into more
homogeneous domains~\cite{Tokura.2006ff}. Such simplifications are often
necessitated by limitations in the experimental methods available to probe more
complex forms of structural inhomogeneity that fall between these two
extremes~\cite{Dagotto.2005ip}.

Inhomogeneity is most frequently characterized by bulk local probes, such as
NMR, µSR or XAFS, which provide only indirect information on the length scales
of structural correlations. However, scanning tunneling microscopy (STM)
measurements have shown that, at least on the surface, the interplay between
electronic and structural inhomogeneity can occur on a range of length scales,
leading to complex short-range order resulting from phase competition and
competing interactions. Pair distribution function (PDF) analysis of both
neutron and x-ray powder diffraction data is a widely used bulk probe of
crystalline disorder~\cite{Billinge.2004uu, Keen.2020}, although some structural
fluctuations, such as transverse displacements, are obscured by spherical
averaging and long-range correlations are hard to separate from the average
structure. Nevertheless, the relative ease of measurement and the sophistication
of PDF analysis software has made this a powerful and popular tool for both
identifying and characterizing short-range order at length scales of a few tens
of {\AA}ngstroms~\cite{Farrow.2007}. 

Single-crystal diffuse scattering, with both neutrons and x-rays, can overcome
many of the limitations of these techniques~\cite{Welberry.2022, Nield.2001wg}.
The method is sensitive to three-dimensional structural correlations over length
scales of 5 to 200 \AA\ or more, and it provides information on both local
atomic relaxations around point defects and the ways defects can self-organize
on the nanoscale into more complex short-range order. In the past, it was
challenging to measure diffuse scattering, because the signal is generally
several orders of magnitude weaker than Bragg scattering and spread over many
Brillouin zones. However, instrumental developments over the past ten to fifteen
years now make possible the collection of large contiguous volumes of scattering
in reciprocal space, encompassing hundreds and often thousands of Brillouin
zones, on time scales ranging from a few minutes, \textit{e.g.}, with
synchrotron x-rays at the Advanced Photon Source~\cite{Krogstad.2019tc}, to a
few hours with neutrons, \textit{e.g.}, on the \textit{CORELLI} diffractometer
at the Spallation Neutron Source~\cite{Ye.2018bc}. Such speeds enable diffuse
scattering data to be collected as a function of temperature and composition,
allowing the evolution of structural fluctuations to be tracked across entire
phase diagrams in a matter of days.

As such experiments have become increasingly routine, the challenge has shifted
to modeling the large three-dimensional (3D) data volumes. Software that
performs atomistic simulations of diffuse scattering, such as
\textit{DISCUS}~\cite{Proffen.1997fc}, provides the necessary computational
framework to calculate the structure factor $S(\mathbf{Q})$, but optimizing the
conditional interatomic vector probabilities that underlie such models is
difficult to perform reliably. However, the data sets are now comprehensive
enough to enable new modes of analysis, such as the 3D-$\Delta$PDF method
pioneered at ETH Z\"urich~\cite{Weber.2012en, Osborn.2023}, which are much
simpler to interpret, even without large-box simulations. 3D-$\Delta$PDF
analysis converts broad reciprocal space intensity distributions into a discrete
set of peaks in real space that represent only those interatomic vector
probabilities that differ from the average crystalline
structure~\cite{Krogstad.2019tc}. Remarkably, even without atomistic modeling,
3D-$\Delta$PDF maps can reveal the length scales over which such structural
deviations are correlated and therefore provide a novel method of extracting
critical exponents in real space both above \textit{and} below structural phase
transitions~\cite{Upreti.2022}.

In this review, we will discuss a number of recent examples in which
comprehensive single-crystal diffuse scattering measurements have transformed
our understanding of quantum materials. These examples cover a broad range of
correlated electron behavior over large length-scale ranges. The work has
provided unique insights into metal-insulator transitions, unconventional
superconductivity, Goldstone mode fluctuations, the effect of disorder on
charge-density waves, and many other phenomena.  Diffuse magnetic neutron
scattering, which also implicitly probes electronic fluctuations, is outside the
scope of this review. The review is divided into five sections. The first
discusses the computational approaches required to interpret the large data sets
now collected on a routine basis. The second section covers the interplay
between electronic and structural fluctuations, and their respective length
scales, at electronic phase transitions. The third describes experiments probing
inhomogeneity in superconducting oxides. Then, recent experiments that probe
the role of extended defects on both unconventional superconductivity and
spin-liquid behavior will be described, before a final section discussing the
outlook for such investigations in the future. 

\section*{Computational Approaches to Diffuse Scattering}

The instrumental advances that have enabled large volumes of $S(\mathbf{Q})$ to
be collected on a routine basis require the development of more efficient
computational tools to process the data and facilitate scientific
interpretation. Even after data reduction, a single dataset can exceed 10~GB in
size, making conventional approaches to data analysis impractical or prone to
selection bias. The problem is only exacerbated by the ever increasing speed of
data collection, particularly at synchrotron x-ray sources, where measurements
typically take less than half an hour, so that they are often repeated at many
temperatures and for a range of sample compositions. It is clear that novel
computational methods are necessary to ensure that more than a small fraction of
the data is utilized.

Two approaches have been developed over the past decade to address this
challenge, and most of the investigations described in this review have utilized
one or both of them. The first is to use machine learning (ML) to identify
significant features in the data by their distinctive temperature dependences.
The second approach is to transform entire data sets into real-space maps of
interatomic vector probabilities, which are usually much easier to interpret
than the original intensity distributions measured in in reciprocal space. We
briefly describe here both methods and then discuss in subsequent sections the
science that they have enabled.

\subsection*{Unsupervised Machine Learning}

\begin{figure}[!t]
  \includegraphics[width=\columnwidth]{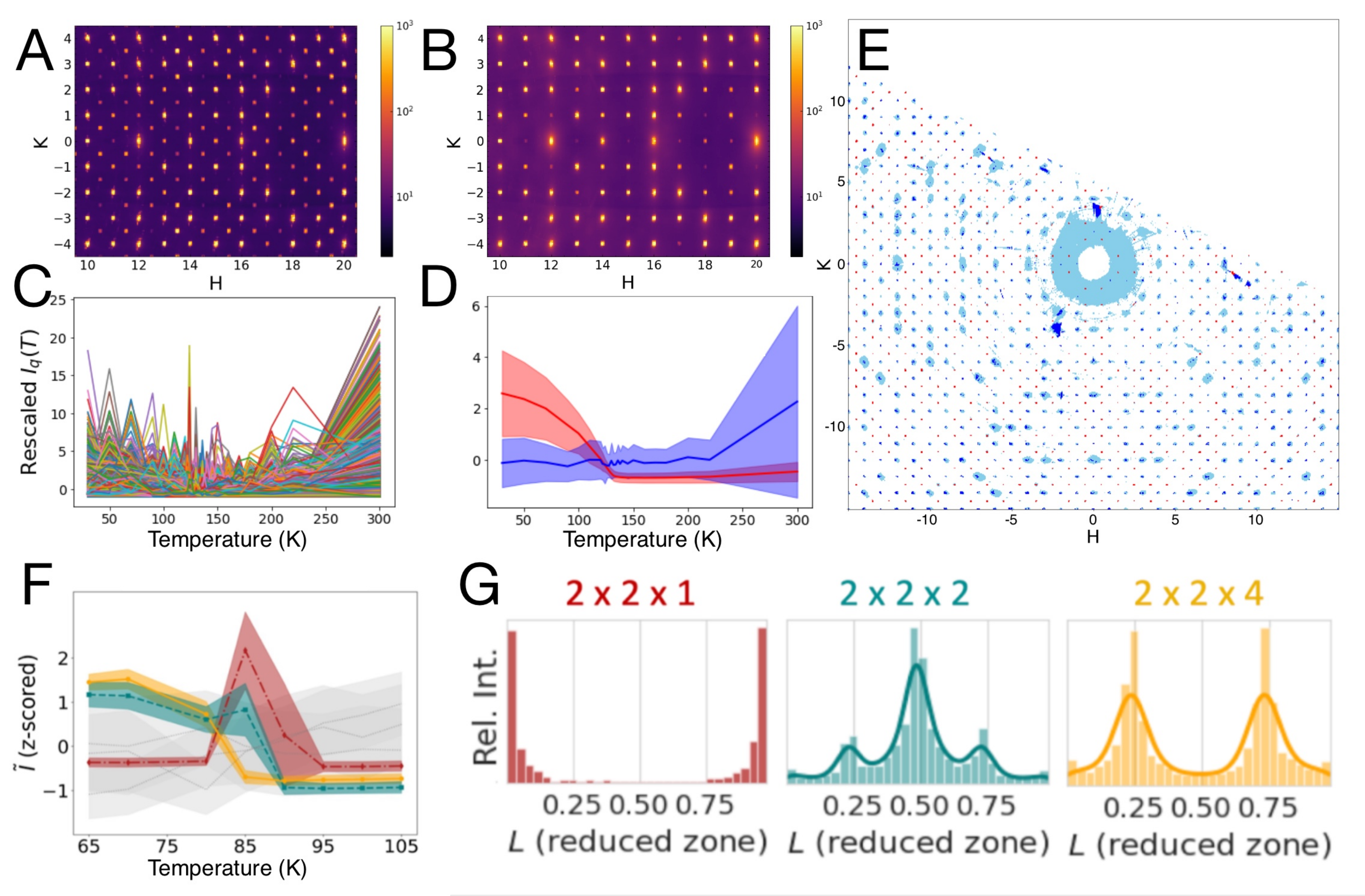} 
  \caption{Examples of the use of \textit{X-TEC} to analyze a billion voxels
  measured in reciprocal space as a function of temperature and cluster them
  according to their distinctive temperature
  dependences~\cite{Venderley.2022}.\@ In Sr$_3$Rh$_4$Sn$_{13}$, data at 30~K
  (\textbf{A}) reveal superlattice peaks, at a relative wavevector of
  (\textonehalf,\textonehalf,0) and symmetric equivalents, which are absent at
  200~K (\textbf{B}).\@ (\textbf{C}) By analyzing the temperature dependence of
  all the voxels, (\textbf{D}) \textit{X-TEC} identified two clusters separating
  superlattice peaks showing typical order-parameter behavior below T$_c$ =
  135~K from other Bragg peaks.\@ (\textbf{E}) A map of these clusters in
  reciprocal space reveal their respective locations.\@ (\textbf{F}) Progression
  of three CDW order parameters in CsV$_3$Sb$_5$ identified by
  \textit{X-TEC}~\cite{Kautzsch.2023}.\@ (\textbf{G}) The momentum distribution
  in the out-of-plane ($L$) axis for each cluster.\label{XTEC}}
\end{figure}

The ability to measure a large number of data sets as a function of a parametric
variable, such as temperature, allows for novel ways to interrogate the data.
For example, unsupervised machine learning algorithms that group voxels within
$S(\mathbf{Q})$ volumes into clusters that share similar temperature dependences
have been implemented in a program called \textit{X-TEC} (\textit{X-ray
diffraction TEmperature Clustering})~\cite{Venderley.2022}. This is a
particularly powerful way to search for a set of superlattice peaks that are
absent above a structural phase transition, but grow with decreasing temperature
with common critical exponents (Fig.\ \ref{XTEC}). This method is not confined
to identifying peaks associated with long-range order, but can also be used to
identify distinct diffuse scattering contributions. The temperature dependence
of each cluster results from the physical origin of the scattering at the
respective wavevectors, so thermal diffuse scattering will typically increase
linearly with temperature, whereas structural diffuse scattering from quenched
defects is often temperature-independent. The temperature dependence of other
diffuse scattering contributions may reflect the growth of emergent order or of
low-frequency critical fluctuations. 

This method was utilized to characterize the sequence of charge-density-wave
(CDW) transitions in the kagome metals CsV$_3$Sb$_5$ and
ScV$_6$Sn$_6$~\cite{Kautzsch.2023, Pokharel.2023}, but it can also be effective
in analyzing diffuse contributions, such as Goldstone modes and short-range CDW
fluctuations above the transition~\cite{Venderley.2022, Pokharel.2023,
Mallayya.2024}, as will be discussed later in this review. Finally, the method
can be extended to provide statistical analyses of scattering associated with
each cluster. For example, it was used to analyze the \textbf{Q}-dependence of
the CDW peak spread to provide evidence of Bragg glass correlations in
Pd$_x$ErTe$_3$~\cite{Mallayya.2024}.

\subsection*{3D-$\Delta$PDF Maps of Structural Correlations}

Another way to ensure that \textit{all} the data collected in single crystal
diffuse scattering experiments are utilized is to transform the reciprocal space
data to real space. When performed over a sufficiently large contiguous
scattering volume, such transforms produce 3D PDFs, \textit{i.e.}, maps of
interatomic vector probabilities summed over all the atomic sites. These are
equivalent to Patterson maps, which have long been used in crystallography as a
tool for solving crystal structures~\cite{Patterson.1934}. When both Bragg peaks
and diffuse scattering are included in the transforms, the PDF maps contain
peaks at interatomic vectors from both the average structure and local
deviations from the average.

\begin{figure}[!b]
  \includegraphics[width=\columnwidth]{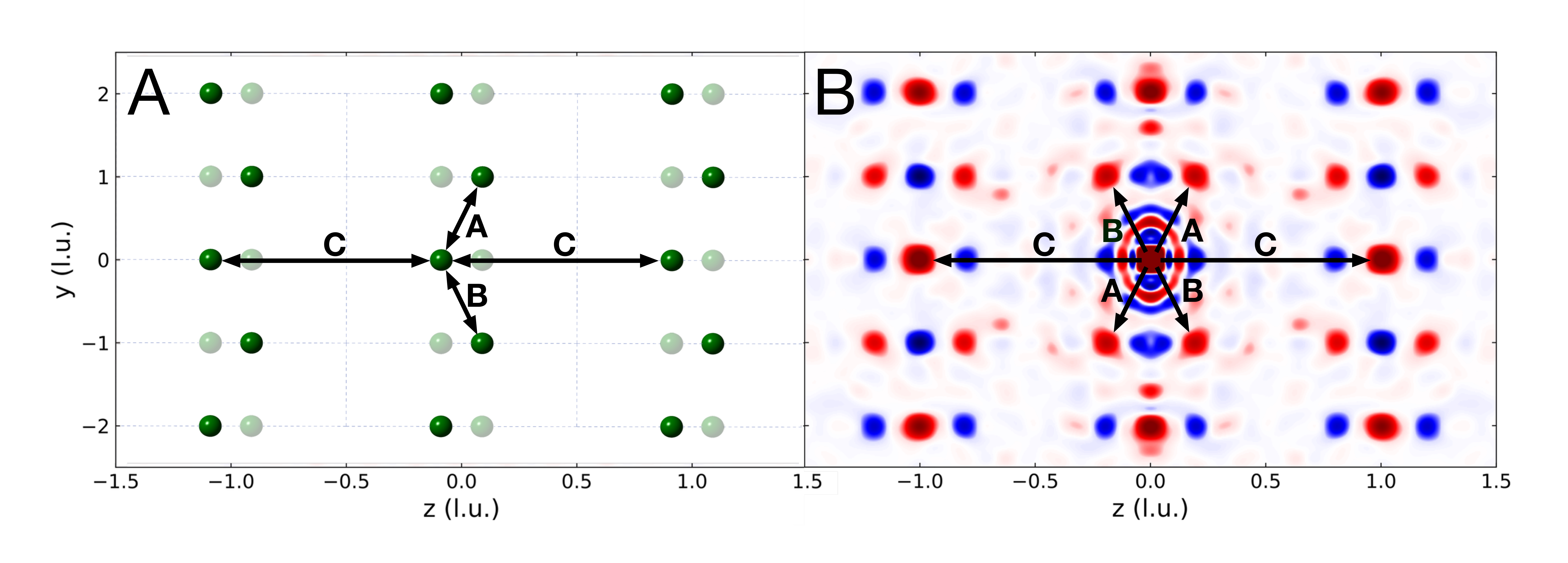}
  \caption{(\textbf{A}) The sodium ions in Na$_x$V$_2$O$_5$ occupy sites on
  two-leg ladders with approximately 50\% of the sites occupied in a zigzag
  configuration.\@  (\textbf{B}) This is evident from the 3D-$\Delta$PDF maps,
  which consist of peaks at real-space interatomic vectors connecting sites that
  are occupied with a greater probability (red) or lesser probability (blue)
  than in the average structure~\cite{Krogstad.2019tc}. The correspondance
  between real space and 3D-$\Delta$PDF maps are illustrated by three
  interatomic vectors labelled A, B, and C. Since the vector origin can be on
  sites on either leg, a two-leg ladder produces a three-leg ladder in the
  3D-$\Delta$PDF\label{V2O5}.}
\end{figure}    

One-dimensional (1D) PDFs derived from powder diffraction data are regularly
used to identify deviations from the average crystallographic structure.
However, in 1D, it is not possible to separate contributions from the average
crystal structure and local deviations without sophisticated modeling, because
they overlap when spherically averaged. However, Thomas Weber and colleagues at
the ETH Z\"urich realized that this is not the case for 3D-PDFs, since the Bragg
peaks are highly localized in reciprocal space. If the Bragg peaks are removed
from S(\textbf{Q}) and replaced by interpolations of the surrounding diffuse
scattering, a procedure known as ``punch-and-fill'', the resulting PDFs only
contain peaks at interatomic vectors whose probabilities differ from the
average. These difference pair-distribution-functions are now known as
3D-$\Delta$PDF maps~\cite{Weber.2012en, Krogstad.2019tc, Osborn.2023}.

Representing the experimental results as 3D-$\Delta$PDF maps produces an
effective dimensional reduction of the raw data, with broad distributions of
diffuse intensity transformed into discrete interatomic vector probabilities. As
Richard Welberry has pointed out, the 3D-$\Delta$PDF peak intensities ``are
simply related to the Warren-Cowley short-range order parameters that have
frequently been used to parametrize diffuse scattering
models''~\cite{Welberry.2022}. These Warren-Cowley parameters, which are the
conditional probabilities for each set of interatomic pairs, effectively
represent all that can be determined about local structural correlations from
S(\textbf{Q}). 3D-$\Delta$PDF maps are usually intuitive to interpret without
computationally expensive modeling and, at the very least, provide a guide to
the appropriate disorder models. For example, in sodium-intercalated V$_2$O$_5$,
it was possible to determine that sodium ions on two-leg ladders only occupy
next-nearest-neighbor sites, forming a zig-zag pattern, without requiring any
further modeling (Fig.~\ref{V2O5})~\cite{Krogstad.2019tc}.

Locally correlated displacements can also be seen via 3D-$\Delta$PDF.\@ A good
example of this is the correlated dipole system PbTe, whose simple rock salt
crystal structure hosts local dipoles formed by off-centered Pb
atoms~\cite{Sangiorgio.2018gj, Holm.2020}. These Pb displacements are locally
correlated only over several unit cells, leaving the average rock salt structure
unchanged while producing broad diffuse scattering rods. The 3D-$\Delta$PDF of
this diffuse scattering shows the distinct quadrupolar signature of a positive
displacement correlation as well as the length scale of the correlation.
3D-$\Delta$PDF analysis is also capable of imaging more complex displacement
correlations, including those coupled to local occupation, as well-demonstrated
in studies of bixbyite~\cite{Stockler.2022}. While this system is better-known
as a case study of diffuse magnetic scattering~\cite{Roth.2019}, it also hosts
complex structural short-range order. Using both reciprocal space and
3D-$\Delta$PDF maps, the authors were able to assemble a detailed model of
stacking faults, intergrowths, and relaxations consistent with the long-range
structure, definitively determining the local structure and showing the
applicability of 3D-$\Delta$PDF beyond the simplest crystal systems.

\section*{Structural Correlations \textit{vs} Electronic Correlations}

Changes in the electronic structure are usually accompanied by changes in the
atomic structure and \textit{vice versa}, but the respective length scales of
electronic and structural correlations have historically been difficult to
ascertain. In some cases, the coupling of the two order parameters is so strong
that both undergo simultaneous, possibly first-order phase transitions, but in
other cases, the interplay is more subtle. Diffuse scattering provides a direct
probe of the structural response to changes in the electronic structure, whether
short-range or long-range. We will describe examples in which structural
correlations profoundly alter our understanding of the physics that underlies
the electronic transitions observed in transport and spectroscopic measurements.

\subsection*{Metal-Insulator Transitions}

\begin{figure}[!b]
  \centering
  \includegraphics[width=0.8\columnwidth]{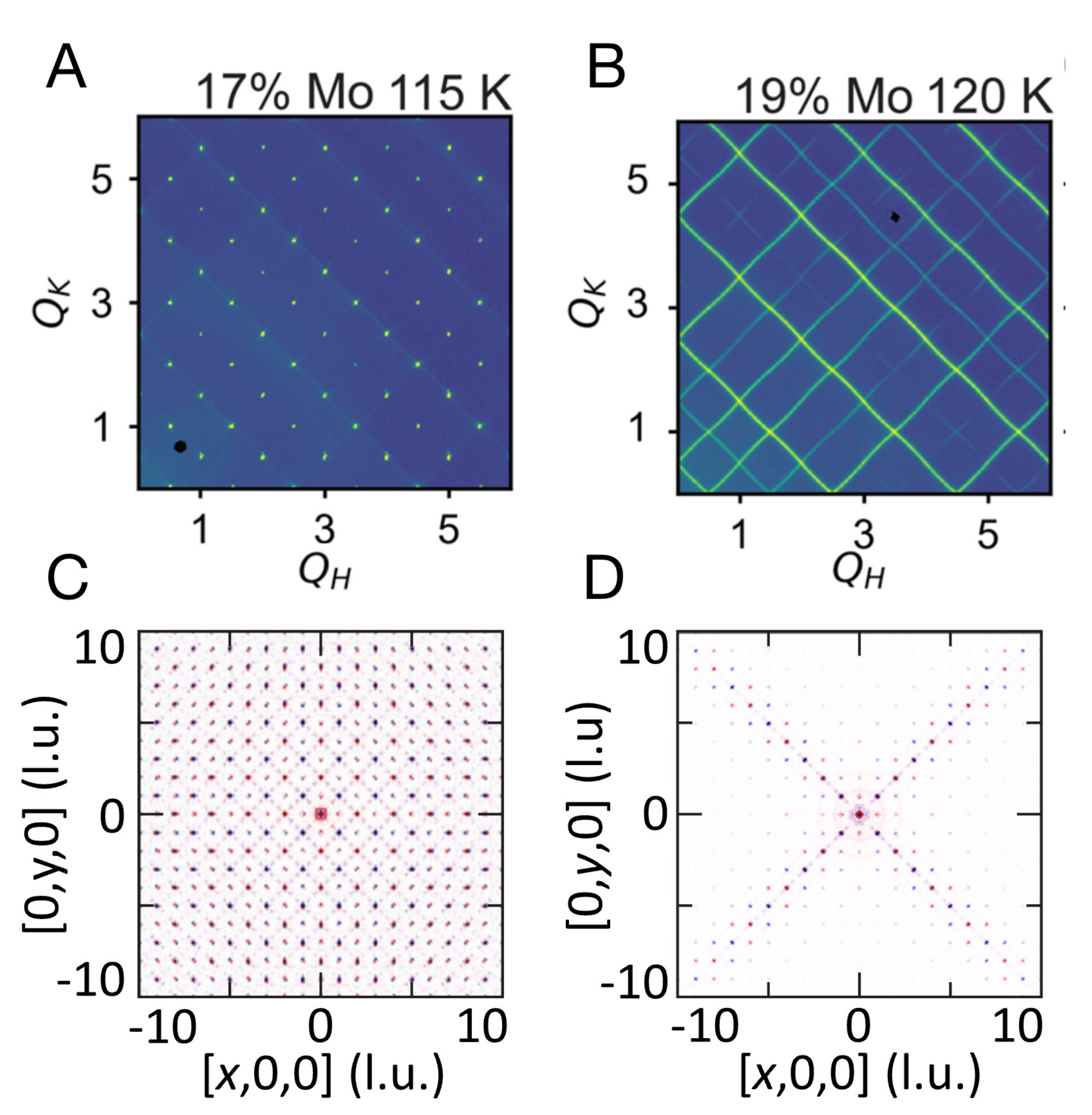}
  \caption{Diffuse scattering from V$_{1-x}$Mo$_x$O$_2$ below the first-order
  electronic phase transition, with (\textbf{A}) $x=0.17$ and (\textbf{B})
  0.19 in the $L=$ \textonehalf\ plane. At $x=0.17$, superlattice peaks
  consistent with the $M_1$ phase appear at the MIT, but at higher $x$-values,
  these are replaced by rods of scattering, indicating short-range
  two-dimensional correlations characteristic of the $M_2$ phase. This is
  confirmed by the respective 3D-$\Delta$PDF maps, which show (\textbf{C})
  long-range $M_1$ order at $x=0.17$, but (\textbf{D}) two-dimensional
  correlations [110] directions at
  $x=0.19$~\cite{Davenport.2019}\label{MIT}.}
\end{figure}    

Metal-insulator transitions (MITs) are often concomitant with structural phase
transitions, whether they are driven by local Mott physics or by a Peierls
distortion~\cite{Imada.1998ti, Georgescu.2022}. We will discuss the specific
example of VO$_2$. On the one hand, there are reports of an apparent suppression
of any structural response to the MIT in epitaxial films of
VO$_2$~\cite{Yang.2016, Ji.2021}. In contrast, in bulk VO$_2$, the strongly
first-order MIT occurs at 340 K, with a transition from a high-temperature
tetragonal rutile structure to the monoclinic $M_1$ phase, in which the vanadium
ions dimerize along buckled $c$-axis chains~\cite{Goodenough.1971dz}. However,
the substitution of other transition metals for vanadium can induce the related
$M_2$ phase, in which only half the vanadium ions dimerize~\cite{Pouget.1974fv},
which indicates that the $M_1$ phase consists of a superposition of the two
equivalent $M_2$ phases along orthogonal [110] directions~\cite{Rice.1994fr}.  

Molybdenum substitution suppresses the MIT from 340 K to about 150 K in
V$_{1-x}$Mo$_x$O$_2$ with $x=0.19$~\cite{Holman.2009ft} and weakens the strength
of the electronic transition. Nevertheless, the transition remains first-order,
so it was a surprise that diffuse scattering measurements showed a complete
collapse of the structural phase transition~\cite{Davenport.2019}. At $x=0.17$,
superlattice peaks in the $L=$ \textonehalf\ plane indicate the development of a
long-range $M_1$ structure below the MIT\@. However, at higher doping, the
superlattice peaks are replaced by wavy rods that indicate purely
two-dimensional correlations persisting over length scales of 50 \AA\ or less
along $\langle 110 \rangle$ directions (Fig.~\ref{MIT}).\@ The waviness results
from weak correlations transverse to these$\langle 110 \rangle$ directions.

3D-$\Delta$PDF maps confirm that the local distortions at $x=0.19$ correspond to
the $M_2$ structure, in which only half the vanadium ions form dimer pairs. This
behavior is also seen in niobium-doped VO$_2$~\cite{Chhetri.2022}. Structurally,
the short-range character of the correlations results from a geometric
frustration of the phase of the lattice buckling in neighboring planes.
Electronically, this result demonstrates that only short-range lattice
relaxations are required to stabilize long-range modifications to the electronic
bands~\cite{Georgescu.2022}, and this is the likely explanation for the absence
of structural changes at the MIT of thin films of
VO$_2$~\cite{Yang.2016,Ji.2021}.

\subsection*{Order-Disorder Transitions}

For over sixty years, second-order structural phase transitions have been
discussed in terms of two limiting categories, as either ``displacive'' or
``order-disorder'' transitions~\cite{Bussmann-Holder.2006}. In the former, the
amplitude of the time-averaged local distortions that define the low-temperature
phase fall to zero at $T_c$, whereas in the latter category, local distortions
persist above $T_c$ but only become phase-coherent at the transition. This
distinction has important implications for the origin of the phase transition
and the nature of the associated changes in electronic structure. If the
transition is order-disorder, then the electronic excitations could become
incoherent above $T_c$ because of the persistence of quasi-static disorder in
the high-temperature phase.

\begin{figure}[!t]
  \includegraphics[width=\columnwidth]{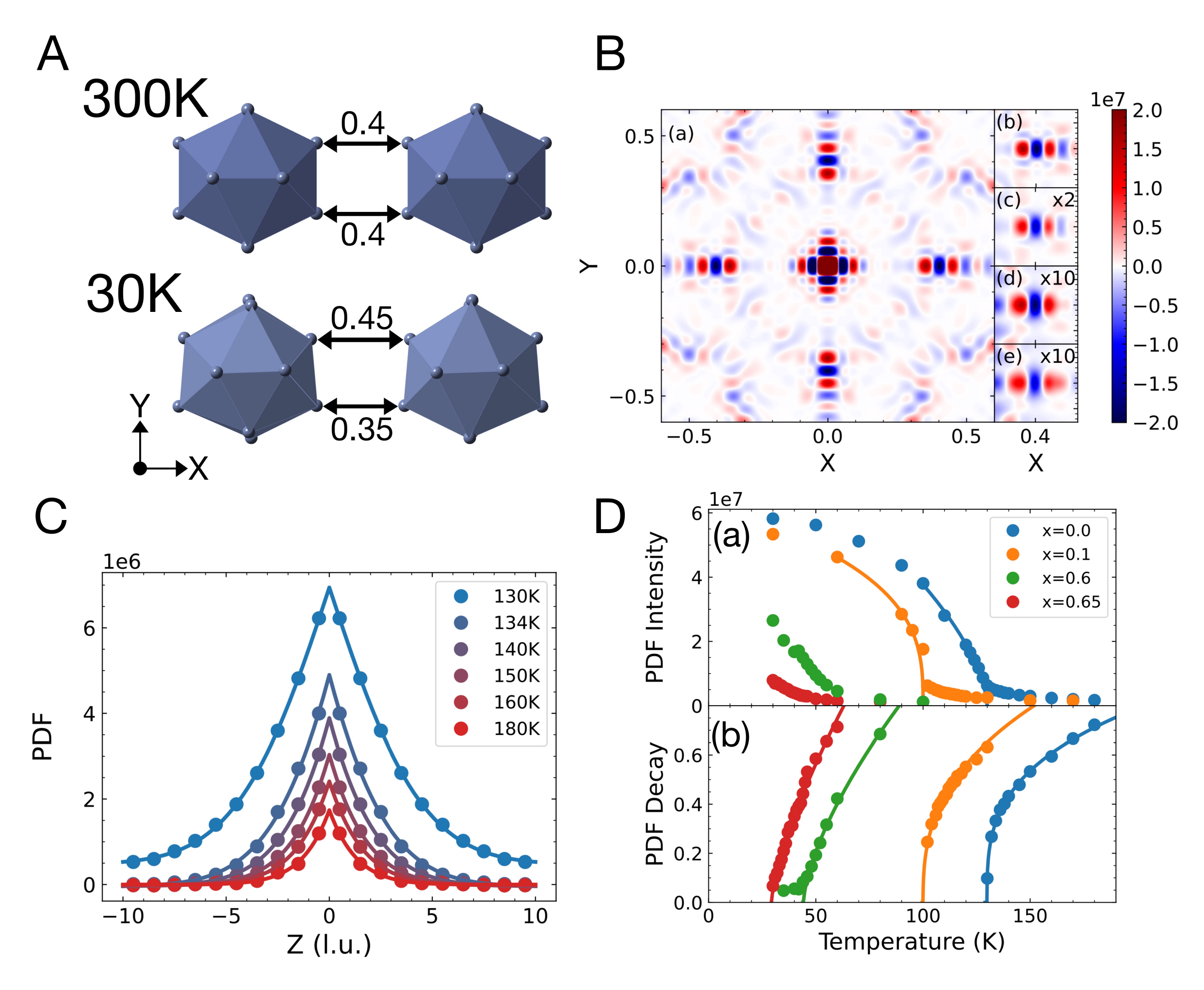} 
  \caption{(\textbf{A}) Neighboring tin icosahedra in Sr$_3$Rh$_4$Sn$_{13}$
  above and below the structural phase transition.\@ (\textbf{B}) 3D-$\Delta$PDF
  maps of the interatomic vector probabilities in the X-Y plane (Z=0). Positive
  values (red) at $\pm$X, $\pm$Y = 0.35 and 0.45 show that the tin ion
  displacements from their high-temperature average positions (blue) at $\pm$X,
  $\pm$Y = 0.4 do not change with temperature, even above T$_c$.\@ (a) 30~K, (b)
  100~K, (c) 120~K, (d) 130~K, and (e) 150 K.\@ (\textbf{C}) Fits of PDF peak
  intensities to an exponential decay showing the correlation lengths in real
  space.\@ (\textbf{D}) Critical scaling curves for
  (Sr$_{1-x}$Ca$_x$)$_3$Rh$_4$Sn$_{13}$ at $x =$ 0, 0.1, 0.6, and 0.65 showing
  the temperature dependence of (a) the order parameter and (b) the inverse
  correlation lengths~\cite{Upreti.2022}.\label{order-disorder}}
\end{figure}      

Diffuse scattering measurements on the quasi-skutterudite Sr$_3$Rh$_4$Sn$_{13}$
provide strong evidence of just such a scenario~\cite{Upreti.2022}
(Fig.~\ref{order-disorder}). This compound undergoes a structural phase
transition at 135 K in which neighboring tin icosahedra counter-rotate, to a
first approximation (Fig.~\ref{order-disorder}A), with the emergence of
superlattice peaks at $\mathbf{q}_s$ = (\textonehalf\,\textonehalf\,0) and
equivalent wavevectors. The growth of these peaks with decreasing temperature
could either be due to an increase in the distortion amplitude or an increase in
phase coherence, but conventional analysis cannot distinguish between the two.
However, 3D-$\Delta$PDF transforms of the diffuse scattering data show that the
distortion amplitudes are independent of temperature
(Fig.~\ref{order-disorder}B), from 30 K to at least 200 K, \textit{i.e.}, even
above the transition, the distortions are undiminished. This is only possible
because the 3D-$\Delta$PDF transforms include both the superlattice peaks and
the surrounding diffuse scattering, which allows the real-space variations of
the PDF peak intensities to be used to measure the correlation lengths of
structural fluctuations from 10 to 200 \AA, both above \textit{and} below $T_c$
(Fig.\ \ref{order-disorder}C), and their respective critical exponents to be
determined (Fig.\ \ref{order-disorder}D).

It had been suggested that Sr$_3$Rh$_4$Sn$_{13}$ and related skutterudites were
unusual examples of systems with three-dimensional CDW
order~\cite{Klintberg.2012dt}, although the nature of the charge
disproportionation was never established. However, the 3D-$\Delta$PDF results
call that into question. Instead, optical spectroscopy provides evidence of
pseudogap behavior~\cite{Ban.2017}, which suggests a loss of electronic coherence
caused by quasi-static structural disorder persisting well above the transition
(see also~\cite{Chatterjee.2015cw}).

It is commonly thought that most transitions display a mixture of properties
associated with both displacive and order-disorder
character~\cite{Bussmann-Holder.2006}. However, the 3D-$\Delta$PDF method, which
allows distortion amplitudes to be tracked down to very low temperatures where
thermal activation of structural disorder is minimized, can remove some of the
ambiguities inherent in other classification methods~\cite{Onodera.2004}.

\subsection*{Structural Fluctuations in Spin-Orbit Coupled Systems}

\begin{figure}[!b]
  \includegraphics[width=\columnwidth]{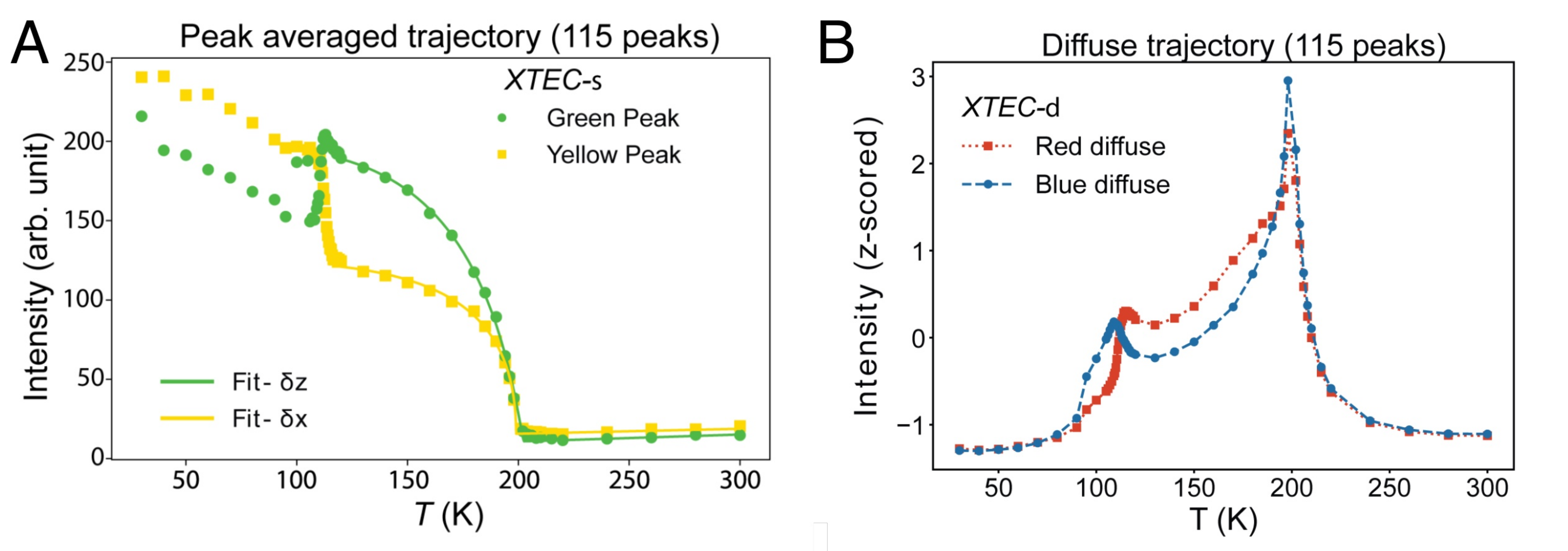} 
  \caption{(\textbf{A}) Temperature dependences of two clusters of superlattice
  peaks (green and yellow) identified by unsupervised machine learning, and
  (\textbf{B}) temperature dependences of diffuse scattering from Goldstone
  mode fluctuations~\cite{Venderley.2022}. The red and blue diffuse clusters
  correspond to the yellow and green superlattice clusters,
  respectively.\label{goldstone}}
\end{figure}

It has been theoretically predicted that metallic systems with strong spin-orbit
coupling will exhibit a variety of novel electronic phases, such as multipolar
nematicity and metallic ferroelectricity~\cite{Fu.2015cl}, but the structural
response in candidate systems is often anomalously small~\cite{Cao.2018be}. One
example is Cd$_2$Re$_2$O$_7$, which is a pyrochlore metal with two structural
phase transitions at 113~K and 200~K that have been attributed to a
parity-breaking nematic transition~\cite{Harter.2017dr, Norman.2020hn}. The
phase transitions produce very small lattice distortions, making conventional
crystallographic analysis extremely challenging~\cite{Yamaura.2002fk,
Castellan.2002kz, Weller.2004gf}. Second-harmonic-generation (SHG) optical
spectroscopy, which is extremely sensitive to parity-breaking phases, seemed to
indicate that the primary order parameter is inconsistent with earlier
diffraction measurements~\cite{Harter.2017dr}, so synchrotron x-ray experiments
initially sought to resolve this question. Use of \textit{X-TEC} identified four
classes of Bragg peaks (two of which are shown in Fig.~\ref{goldstone}A) whose
temperature dependences showed distinct behavior at the two phase transitions.
Selection rules derived from the \textbf{Q}-dependence of these clusters were
associated with the sensitivity of the structure factors to in-plane and
out-of-plane cation distortions.

The earlier structural investigations indicated that the upper transition
involved a lowering of cubic symmetry (space group $Fd\overline{3}m$) to
tetragonal symmetry (space group $I\overline{4}m2$), which is produced by a
condensation of two-component \textit{E$_u$} modes~\cite{Yamaura.2002fk,
Norman.2020hn}. If these components are nearly degenerate, there would be strong
Goldstone mode fluctuations between the two, corresponding to a switch between
in-plane and $z$-axis ionic distortions. Raman scattering had already found some
evidence of the existence of such Goldstone modes, so \textit{X-TEC} was used to
determine their \textbf{Q} dependence by analyzing diffuse scattering around
hundreds of Bragg peaks. Stronger fluctuations were observed around Bragg peaks
that were particularly sensitive to in-plane distortions
(Fig.~\ref{goldstone}B). This is consistent with a Landau theory analysis, which
predicts fluctuations towards the second \textit{E$_u$} component with
$I4_{1}22$ symmetry, which then condense at the lower first-order
transition~\cite{Venderley.2022}. This is the first time that selection rules for
diffuse scattering have been correlated with those for superlattice reflections,
which provided detailed insights into the mechanisms that drive the two
structural phase transitions.

\subsection*{Bragg Glass Correlations}

Nearly fifty years ago, Imry and Ma predicted that ordered phases with
continuous symmetry would be unstable to arbitrarily weak random fields below
the upper critical dimension of four~\cite{Imry.1975bi}. One example is the
destruction of long-range order when phase fluctuations are pinned by random
defects in an incommensurate CDW.\@ The competition between the disorder
potential and elastic strain was predicted to result in a vestigial nematic
phase with a short-range correlation length~\cite{Fukuyama.1978}. However, it
was later pointed out that this prediction is modified when the periodicity of
the phase is taken into account~\cite{Nattermann.1989}. Instead, strong phase
fluctuations are predicted to result in an unusual form of
quasi-long-range-order, in which there is an algebraic decay of structural
correlations~\cite{Giamarchi.1994, Giamarchi.1995}. This was termed a Bragg
glass, a completely new phase that has been extremely difficult to observe
because of the exacting resolution required to distinguish it from true
long-range order. 

STM studies of incommensurate CDW compounds have provided evidence of Bragg
glass behavior through an analysis of topological defects in the presence of
disorder~\cite{Okamoto.2015vo, Fang.2019ha}, but the first bulk-probe evidence
was only recently provided by a novel ML analysis of diffuse x-ray scattering
data. The suppression of CDW order in palladium-intercalated ErTe$_3$ has been
extensively studied by scattering, transport, as well as STM
measurements~\cite{Straquadine.2019ju, Fang.2019ha}. In pure ErTe$_3$, there are
two CDW phases with transition temperatures of 260 K and 135 K, resulting from
orthogonal modulations of the tellurium square-planar
nets~\cite{Straquadine.2019ju}. The lower transition is rapidly suppressed by an
intercalation of less than 1\% palladium, but evidence of CDW peaks from the
upper transition persist up to 3\% intercalation. However, the momentum
resolution of x-ray diffraction is insufficient to determine if these peaks
display the power-law tails characteristic of a Bragg glass.

To overcome this limitation, \textit{X-TEC} was modified to investigate the
linewidths of the broadened CDW peaks above the apparent CDW
transitions~\cite{Mallayya.2024}. The theoretical basis of this analysis is that
peak broadening due to phase fluctuations is $Q$-independent and so can be
distinguished from the quadratic $Q$-dependence of displacement fluctuations. By
a novel analysis of the peak spread of thousands of CDW peaks, it proved
possible to establish that the correlation lengths associated with phase
fluctuations diverge at a nonzero temperature, whereas a vestigial nematic
should remain short-range at all temperatures. The inferred phase diagram of the
Bragg glass phases is in good agreement with the onset of in-plane anisotropy in
transport measurements~\cite{Straquadine.2019ju}.

It is evident from this example that the unsupervised machine learning approach
implemented by \textit{X-TEC} can be adapted to a variety of different problems
whenever there are sufficient data to generate robust statistical analyses of
both the $Q$- and temperature-dependence of features in reciprocal space. 

\section*{Superconducting Oxides}

Complex oxides that exhibit high-temperature superconductivity have been
intensely studied for more than three decades, including a wide array of
experiments sensitive to structural inhomogeneity. Yet the extent and role of
nanoscale structural correlations in the phenomenology of these materials remains heavily
debated, often due to tremendous difficulties in resolving the many different
kinds of disorder. The lamellar high-$T_c$ cuprates have been most heavily
investigated due to their extraordinary superconducting and normal-state
properties~\cite{Keimer.2015}, yet oxides such as the
bismuthates~\cite{Sleight.2015} also show high-temperature superconductivity and
other electronic ordering tendencies that are not well understood. It has been
long known that complex oxides are prone to various types of inhomogeneity of
both electronic and structural origin~\cite{Dagotto.2005ip}. Perhaps most
importantly, the vast majority of superconducting oxides must be chemically
doped to achieve superconductivity, which necessarily introduces point disorder.
Moreover, the most prominent oxide superconductors have perovskite-derived
structures, which generically exhibit structural instabilities due to atomic
size mismatch. The latter can lead to both long-range lowering of the structural
symmetry and short-range fluctuations embedded in a high-symmetry phase. In
addition, a host of inhomogeneous phases of apparent electronic origin has been found,
including spin and charge density waves, with coherence lengths of only a few
unit cells in some systems. Scattering has played an essential role in the
discovery of long- and short-range density wave order in the cuprates, from the
pioneering neutron work on La-based materials~\cite{Tranquada.1995} to more
recent x-ray scattering experiments in multiple cuprate
families~\cite{Ghiringhelli.2012, Chang.2012, Comin.2014, daSilvaNeto.2015,
Arpaia.2019, Yu.2020}. Indeed, the development of sophisticated resonant soft
x-ray scattering techniques has largely been motivated by studies of charge
density waves in cuprates. Most of these experiments, however, have focused on limited
reciprocal space volumes and tiny signals, and have not provided systematic
insight into different types of bulk inhomogeneity. With the recent development
of high-throughput diffuse scattering instruments, 3D-$\Delta$PDF analysis, and
advanced numerical modeling, this important question is beginning to be
addressed in several model systems.

We highlight two sets of results here: the finding of inversion-breaking atomic
correlations in the prototypical superconducting bismuthate
Ba$_{1-x}$K$_x$BiO$_3$ (BKBO)~\cite{Griffitt.2023} and the detailed
characterization of nanoscale structural correlations in the model cuprate
HgBa$_2$CuO$_{4+\delta}$ (Hg1201)~\cite{Anderson.2024}. These examples
comprehensively showcase the strengths and possibilities of state-of-the-art
diffuse scattering and associated analysis, indicate the presence of unexpected
interactions between the local structure and electronic degrees of freedom, and
provide the foundation for a broad range of further investigations.

\subsection*{Bismuthates}

Superconductivity in the bismuthates was discovered nearly five decades
ago~\cite{Sleight.2015}, and the BKBO family shows a maximum $T_c$ above 30 K,
similar to the La-based cuprates~\cite{Cava.1988}. Yet bismuthate research has
been somewhat overshadowed by the cuprates, and major questions pertaining to
the doping-temperature phase diagram and superconducting pairing mechanism
remain open. Although their average structure is close to a simple cubic
perovskite with Bi-O octahedra, the bismuthates display significant structural
and electronic complexity~\cite{Sleight.2015, Pei.1990}. The stoichiometric
parent compound, BaBiO$_3$, is an insulator with pronounced charge
disproportionation: the local charge periodically changes from one Bi-O
octahedron to the next, in what can be viewed as a commensurate CDW.\@ Upon
substitutional doping, either via Bi$\rightarrow$Pb or Ba$\rightarrow$K, the
long-range CDW order disappears, and a metallic/superconducting phase emerges at
sufficiently high doping levels. It has long been speculated that short-range
CDW correlations survive deep into the metallic phase and play an important role
in the superconducting pairing mechanism~\cite{Jurczek.1986, Jiang.2021}.
Alternatively, the bismuthates have been proposed to be conventional
electron-phonon superconductors, with a large electronic coupling to optical
phonon branches that involve oxygen~\cite{Yin.2013}. Since diffuse scattering is
sensitive to short-range CDW correlations, this pivotal conundrum can be
resolved through studies of the local structure. Notably, the Bi-Pb system is
more complicated compared to Ba-K (BKBO), due to the presence of metastable
structural variants~\cite{Sleight.2015}. This leads to interesting mesoscale
structures that might be easily tunable with strain, but it also makes this
bismuthate family less suitable for systematic diffuse scattering studies. BKBO,
in contrast, is nearly ideal: the simple average structure and small unit cell
enable both fruitful 3D-$\Delta$PDF analysis and quantitative modeling.

\begin{figure}[!b]
  \includegraphics[width=\columnwidth]{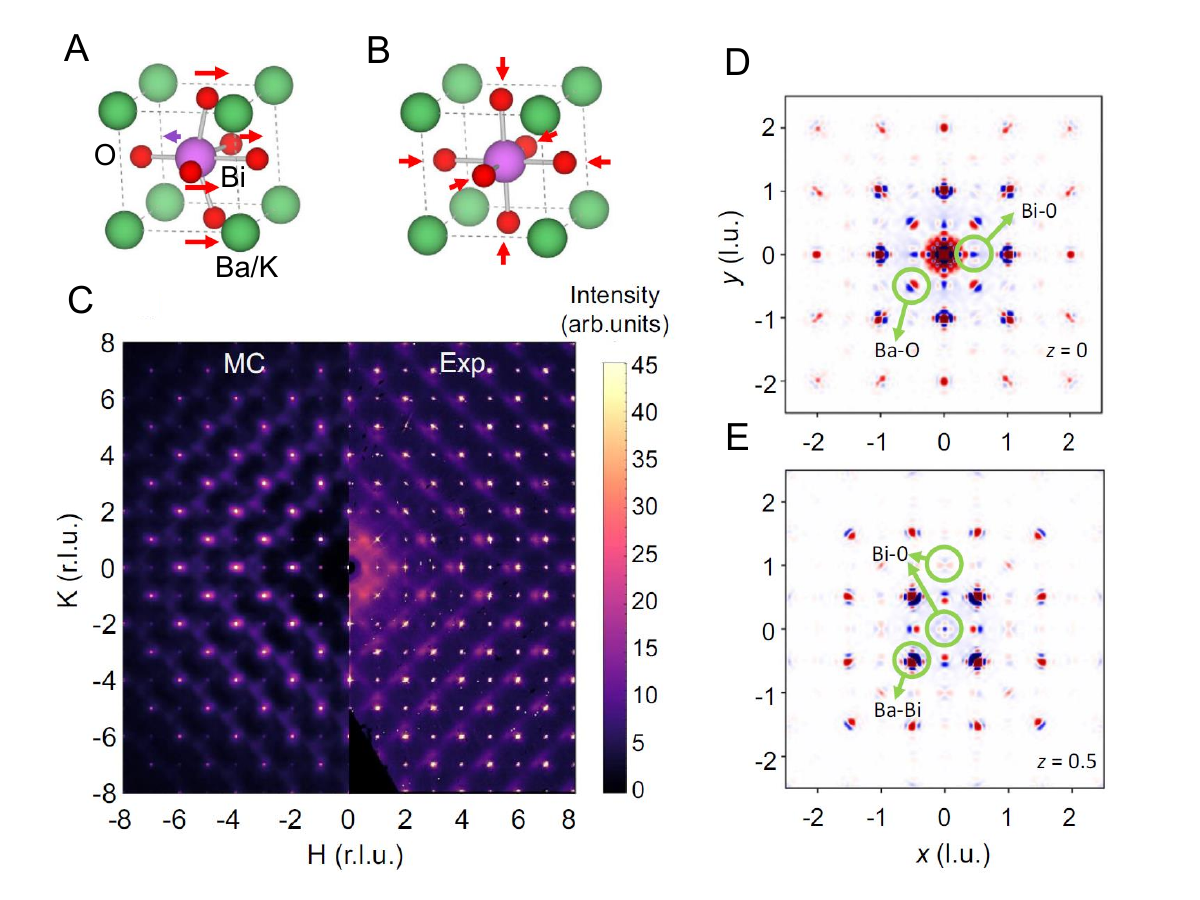} 
  \caption{Local structural correlations in superconducting
  Ba$_{1-x}$K$_x$BiO$_3$~\cite{Griffitt.2023}.\@ (\textbf{A}, \textbf{B}) Two
  characteristic distortions of Bi-O octahedra: an inversion-breaking
  dislacement (\textbf{A}) and a breathing distortion (\textbf{B}). The latter
  is associated with a CDW phase in the parent compound BaBiO$_3$.\@
  (\textbf{C}) X-ray diffuse scattering data for a single crystal of
  Ba$_{0.6}$K$_{0.4}$BiO$_3$ (Exp), compared to classical Monte Carlo modeling
  (MC). A cut with $L = 0$ is shown. The diffuse patterns predominantly
  originate from short-range inversion-breaking distortions of the type seen in
  (\textbf{A}).\@ (\textbf{D}, \textbf{E}) 3D-$\Delta$PDF in the $z=0$ and
  $z=0.5$ planes generated from x-ray scattering data, with the most important
  pair correlations labeled in each panel. No evidence of breathing distortions
  is found, and the opposite signs of Ba-O and Ba-Bi correlations are only
  consistent with an inversion-breaking distortion.\label{bismuthates}}
\end{figure} 

X-ray diffuse scattering measurements on BKBO have yielded two central
results~\cite{Griffitt.2023} (Fig.~\ref{bismuthates}). First, no trace of
short-range CDW correlations is observed in crystals without long-range CDW
order. This includes both insulating and metallic/superconducting BKBO, and it
implies that CDW correlations are likely not relevant for bismuthate
superconductivity. The second important result is the surprising finding of
nanoscale structural correlations that break inversion symmetry
(Fig.~\ref{bismuthates}A). These polar octahedral distortions lead to
characteristic diffuse scattering features (Fig.~\ref{bismuthates}C) that are
much stronger in metallic than in insulating samples. They are also clearly
visible in the 3D-$\Delta$PDF, especially through the opposite signs of the Ba-O
and Ba-Bi correlations (Fig.~\ref{bismuthates}D,E). Moreover, the polar
distortions are seen in classical Monte Carlo simulations based on effective
bond valence sums, similar to previous work on relaxor
ferroelectrics~\cite{Whitfield.2014}. The simulations also provide insight into
the origin of the correlations: an intrinsic tendency toward such octahedral
deformation is amplified by the local charge inhomogeneity introduced by the
Ba-K substitution. In metallic BKBO, electronic screening renders electrostatic
interactions short-ranged, and thus likely both enhances the correlations and
sets their length-scale.

The presence of locally broken inversion symmetry opens up new perspectives in
bismuthate physics. Most importantly, effective Rashba interactions between
conducting electrons and phonons become possible, and might contribute to
superconducting pairing. Moreover, since BKBO turns out to be a locally
noncenstrosymmetric superconductor, the possibility for exotic superconducting
order parameter symmetries arises~\cite{Smidman.2017}. Without inversion, the
usual classification into parity-even and parity-odd superconducting order
parameters is no longer applicable, and mixed-parity states are allowed. In
turn, these might exhibit broken time-reversal symmetry, which could be detected
using complementary local probes such as muon spin rotation. Finally, diffuse
scattering studies of other bismuthate families, as well as related compounds
such as antimonides~\cite{Kim.2022}, should provide further insight into their
similarities and differences, and might help to explain why BKBO shows the
highest $T_c$ values among the bismuthates.

\subsection*{Cuprates}

\begin{figure}[!b]
  \includegraphics[width=\columnwidth]{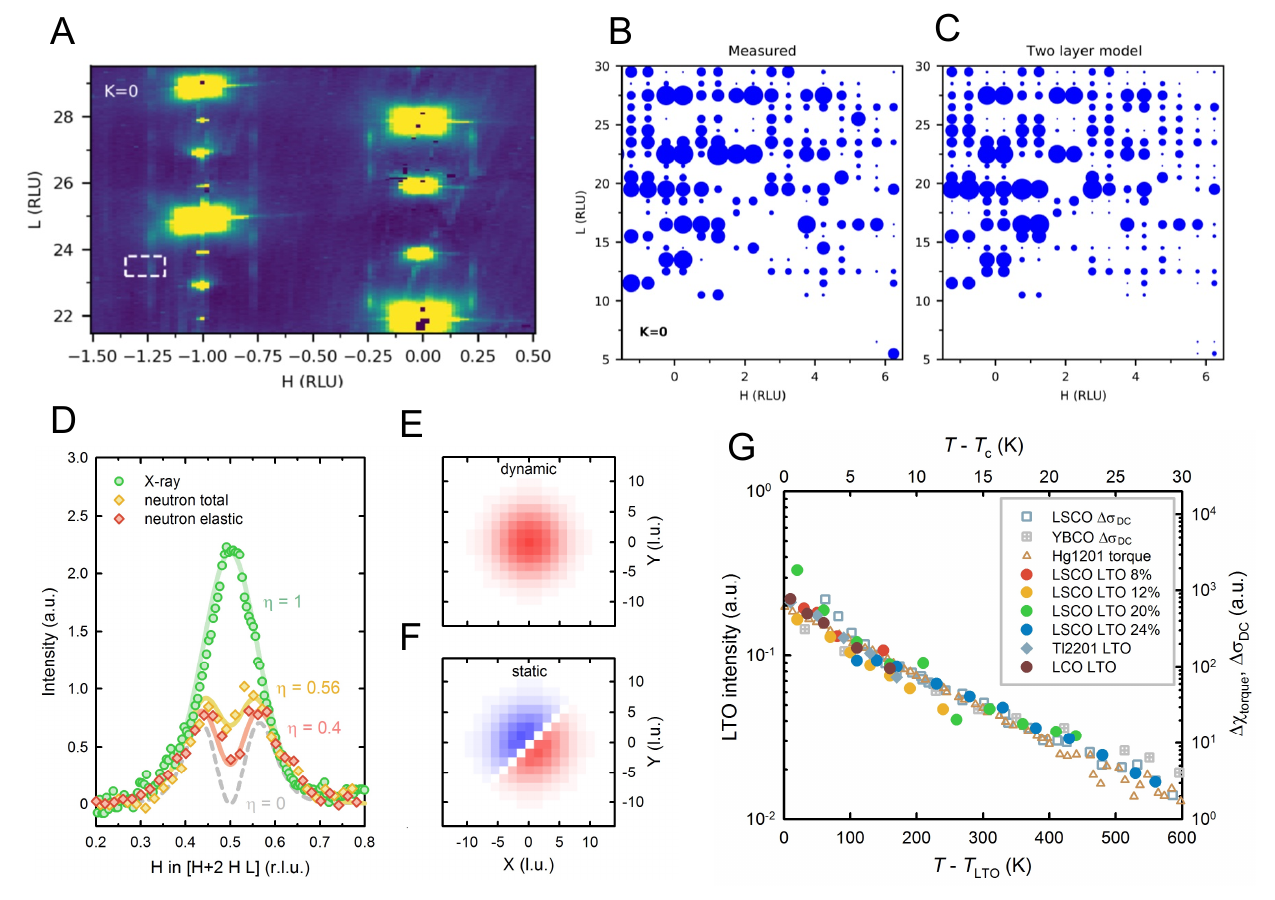} 
  \caption{Diffuse scattering in lanthanum and thallium-based cuprates.\@
  (\textbf{A}) Diffuse scattering in
  La$_{1.875}$Ba$_{0.125}$CuO$_4$~\cite{Sears.2023}.\@ (\textbf{B}) The measured
  CDW peak intensities are compared with (\textbf{C}) those calculated by a
  model of La and Cu modulations in both layers of the crystal structure.\@
  (\textbf{D}) One-dimensional cuts through a half-integer Bragg position in the
  high-temperature tetragonal phase of La$_{2-x}$Sr$_x$CuO$_4$
  (LSCO)~\cite{Pelc.2022}. Neutron scattering measured with CORELLI (red --
  quasielastic scattering; orange -- energy-integrated scattering) shows a split
  diffuse peak, while only a single peak is seen in x-ray scattering. Since
  neutron scattering is sensitive to a significantly smaller energy range than
  x-ray scattering, this stark difference was ascribed to the presence of both
  (\textbf{E}) dynamic and (\textbf{F}) quasistatic orthorhombic fluctuations,
  with the latter showing an antiphase boundary that leads to the observed
  low-energy incommensurability.\@ (\textbf{G}) Exponential scaling of the
  diffuse superstructure intensity above the tetragonal-to-orthorhombic
  transition temperature $T_{LTO}$ for LSCO with several Sr concentrations, as
  well as optimally doped Tl2201, compared to measurements of superconducting
  fluctuations.\label{LTO}}
\end{figure} 

One of the defining features of the lamellar high-$T_c$ cuprates is the
interplay between perovskite-derived copper-oxygen planes and the intervening
ionic rock-salt layers that separate the planes, and this generic structure can
host a wide variety of distortions. Structural and electronic inhomogeneity has
been extensively investigated in the cuprates since the early
days~\cite{Egami.1994}, with numerous prominent STM~\cite{Pan.2001,
Fischer.2007, Du.2020}, NMR~\cite{Singer.2002, Bobroff.2002, Frachet.2020},
x-ray~\cite{Bianconi.1996, Fratini.2010} and neutron
scattering~\cite{Tranquada.1995, Bozin.2000} studies. Yet no consensus has
emerged on the common characteristics and importance of nanoscale correlations,
as different cuprate families exhibit various specific forms of inhomogeneity,
along with doping-related point disorder~\cite{Eisaki.2004}. As noted, the
cuprates harbor short-range charge-density wave order that has been extensively
studied with scattering techniques, including detailed recent diffuse x-ray
scattering work on La$_{2-x}$Ba$_x$CuO$_4$~\cite{Sears.2023}
(Fig.~\ref{LTO}A-C). The Bi-based cuprates, which have been especially favorable
for investigations with surface-sensitive probes, display both long-range and
short-range superstructures~\cite{Castellan.2006,Izquierdo.2006,Poccia.2020}
that can be modified, \textit{e.g.}, by Pb co-doping~\cite{Jakubowicz.2001}.
Another important system, YBa$_2$Cu$_3$O$_{7-\delta}$ (YBCO), shows complex
ordering patterns of oxygen interstitials that cause extensive diffuse
scattering~\cite{Beyers.1989, Strempfer.2004}. Body-centered systems such as the
La-based cuprates La$_{2-x}$Sr$_x$CuO$_4$ (LSCO) and La$_{2-x}$Ba$_x$CuO$_4$
(LBCO) exhibit a series of symmetry-lowering structural transitions due to rigid
rotations of the Cu-O octahedra~\cite{Axe.1994}, and extensive scattering
investigations have shown that the associated short-range fluctuations are
prominent across the temperature-doping phase diagram~\cite{Egami.1994,
Bozin.1998, Wakimoto.2006}. Recent diffuse scattering work established a
universal exponential temperature dependence of short-range orthorhombic
fluctuations in LSCO and the Tl-based system Tl$_2$Ba$_2$CuO$_{6+\delta}$
(Tl2201)~\cite{Pelc.2022}, which closely resembles the superconducting
fluctuation behavior~\cite{Yu.2019, Pelc.2019} (Fig.~\ref{LTO}D-G). This unusual
observation has been interpreted as a signature of rare ordered regions that
appear well above the bulk phase transition temperatures due to some underlying
doping- and family-independent correlated inhomogeneity. If such inhomogeneity
is indeed present, it would have far-reaching consequences for our understanding
of cuprate physics, and diffuse scattering is one of the most versatile tools to
search for it.

\begin{figure}[!b]
  \includegraphics[width=\columnwidth]{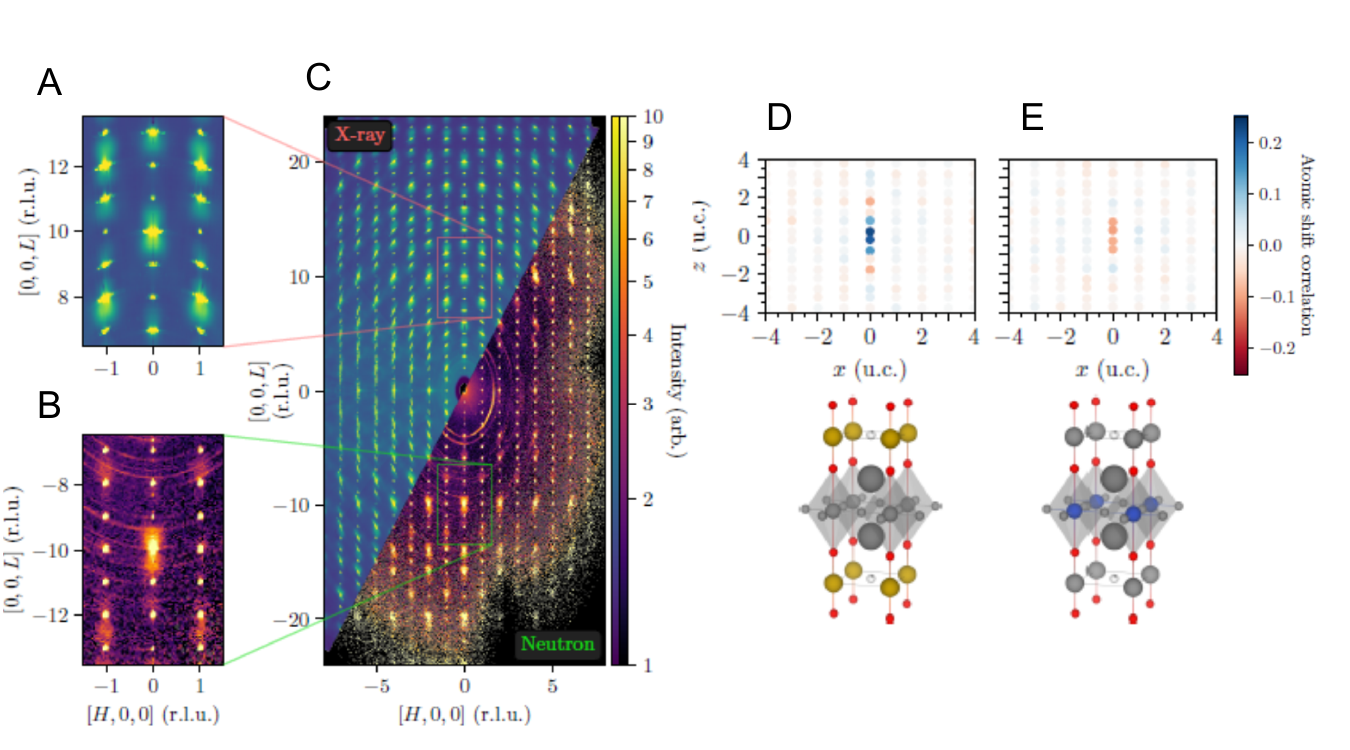} 
  \caption{Local structure of the model cuprate superconductor
  Hg1201~\cite{Anderson.2024}.\@ (\textbf{A})-(\textbf{C}) Combined neutron and
  X-ray scattering data in underdoped Hg1201 crystals that show similar
  lobe-like features in the $HK$ plane, indicative of complex nanoscale atomic
  displacements perpendicular to the Cu-O planes.\@ (\textbf{D}), (\textbf{E})
  Atomic pair correlation functions obtained from reverse Monte Carlo fits to
  the reciprocal space data. Mercury and apical oxygen atoms show strong
  positive correlations (\textbf{D}), while copper and apical oxygen display
  negative short-range correlations (\textbf{E}). This is consistent either with
  the formation of local Hg-O dipoles in the ionic layer, or breathing
  distortions of the Cu-O octahedra.\label{Hg1201}}
\end{figure} 

To this end, HgBa$_2$CuO$_{4+\delta}$ (Hg1201) was chosen as a model system for
detailed diffuse scattering measurements. The main advantage of this compound is
a simple tetragonal average structure and small unit cell, along with an absence
of structural transitions. Moreover, Hg1201 is doped using interstitial oxygen,
which resides relatively far from the qunitessential Cu-O planes and perturbs 
the lattice less severely than substitutional doping. Early diffuse scattering
work~\cite{Couture.2010,Izquierdo.2011, Welberry.2016} uncovered a tendency for
the interstitial oxygen atoms to form chain-like structures in samples with
transition temperatures above about 80 K, yet at lower densities the
interstitials are essentially randomly distributed. Electronically, Hg1201 shows
the highest $T_c$ values of any cuprate with a single Cu-O plane per primitive
cell, as well as quantum oscillations~\cite{Barisic.2013} and negligible
residual resistivities~\cite{Barisic.2013a}, which demonstrates a weak influence
of the interstitials on the Cu-O planes. All this indicates that Hg1201 is one
of the most pristine cuprates and representative of the entire cuprate family.

A combined neutron and x-ray diffuse scattering study of samples with relatively
low interstitial oxygen densities revealed extensive and highly structured
reciprocal-space features in Hg1201, implying that nanoscale structural
correlations are prominent in this material~\cite{Anderson.2024}. It is
immediately clear that there is little diffuse scattering within the $H$-$K$
planes, which implies that the atomic correlations are predominantly in the
out-of-plane direction (Fig.~\ref{Hg1201}). Moreover, the elastic discrimination
enabled by the \textit{CORELLI} instrument at the Spallation Neutron Source
shows that the diffuse features are mostly static, an observation that was
further confirmed in a targeted inelastic neutron scattering experiment. Both
the neutron and x-ray scattering data were of sufficient quality to generate
3D-$\Delta$PDFs, which provide further insight. The characteristic length scales
associated with the correlations turn out to be $\sim 10$ unit cells within the
Cu-O planes and $\sim 3$ unit cells perpendicular to the planes. Interestingly,
both length-scales are comparable to, or larger than the superconducting
coherence lengths. This implies that the pairing is affected by the
inhomogeneity, which could explain the observation of the unusual exponential
fluctuation regime (Fig.\ \ref{LTO}). 

Although Hg1201 has a relatively small unit cell, there is still significant
overlap among different atomic pair vectors in the 3D-$\Delta$PDF, which makes
interpretation challenging. In order to reliably determine the real-space nature
of the atomic displacements, reverse Monte Carlo refinement was employed on both
the x-ray and neutron scattering data in reciprocal space. The large supercells
that are produced numerically enable detailed analysis of atomic correlation
functions, and show that the most important pair correlations are between apical
oxygen and mercury atoms, as well as apical oxygen and in-plane copper. Mercury
and apical oxygen displacements are positively correlated, while the copper and
apical oxygen displacements are anticorrelated, which points to two possible
origins: the formation of local Hg-O dipoles in the ionic layer, or a breathing
distortion of the Cu-O octahedra. While further insights from theory or
\textit{ab initio} modeling might resolve this question, both effects are not
specific to the simple-tetragonal Hg1201 structure and might thus be a generic
property of the cuprates. Diffuse scattering measurements of other cuprate
families will therefore be highly valuable. More broadly, the comprehesive
neutron, x-ray, and numerical work on Hg1201 constitutes both a technical and a
scientific benchmark for structural studies of other quantum materials.

\section*{Extended Defects}

Defects that extend over many unit cells, such as dislocations and stacking
faults, cause specific diffuse scattering signatures, and the new generation of
high-sensitivity instruments enables unprecedented scientific opportunities.
While extended defects are crucial in materials science and metallurgy, they
have been much less studied in the context of quantum materials, although they
can lead to dramatic effects. Dislocations are associated with enormous local
lattice strains, which can cause qualitative changes in the electronic subsystem
in their vicinity. In turn, stacking faults destroy the long-range periodicity
of layered crystal structures, and can thus substantially affect electronically
ordered states. We discuss here two recent prominent examples: self-organized 
dislocation structures induced by plastic deformation in the perovskite oxides 
SrTiO$_3$ (STO) and KTaO$_3$ (KTO), and stacking faults in the van der Waals 
spin-liquid candidate RuCl$_3$.

\subsection*{Plastic Deformation in Oxides}

A straightforward way to introduce dislocations into a material is via
irreversible, plastic deformation. Yet for this to be possible, cracking must
not occur, \textit{i.e.}, the energy of dislocation formation and/or migration
must be sufficiently low. Given that these processes are thermally activated,
most materials become ductile at temperatures that are a sizable fraction of
their melting points. However, some systems display anomalous ductility at much
lower temperatures, with STO a prominent example~\cite{Gumbsch.2001}. STO is
also a well-known quantum material, with low-temperature electronic properties
that have been the subject of debate for six decades~\cite{Collignon.2019,
Gastiasoro.2020}. Pristine STO is a band insulator and incipient ferroelectric:
it is very close to a ferroelectric instability, but does not show long-range
order down to the lowest temperatures. Upon doping with electrons, the material
becomes superconducting at record-low charge-carrier densities and shows
puzzling normal-state transport properties. Importantly, both the
superconductivity and ferroelectricity are extremely sensitive to lattice
strain, which makes STO an ideal model system to study the effects of plastic
deformation on the quantum properties of a material.

\begin{figure}[!b]
  \includegraphics[width=\columnwidth]{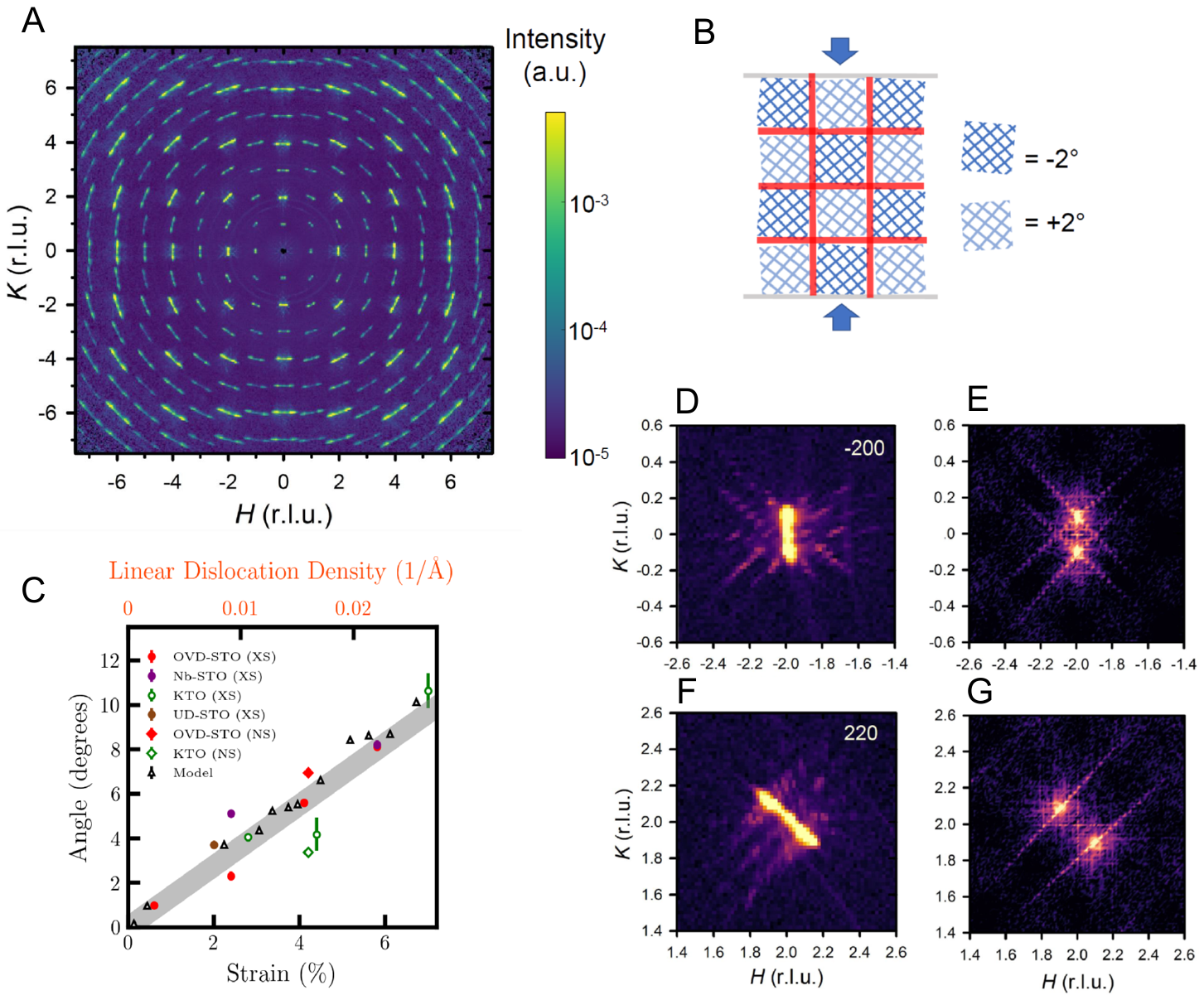} 
  \caption{The structure of plastically deformed strontium titanate.\@
  (\textbf{A}) Neutron scattering data obtained for a STO crystal deformed to
  $\varepsilon = 4.2$\% in compression along $\langle 0 0 1
  \rangle$~\cite{Hameed.2020}. Upon deformation, sharp Bragg peaks transform
  into arcs (asterisms) due to the creation of tilted domains, as shown
  schematically in (\textbf{B}). The dislocations self-organize into walls (red
  lines in (\textbf{B})), where the internal strain is highly concentrated.\@
  (\textbf{C}) Dependence of the asterism angular spread on the strain level for
  STO and KTO crystals~\cite{Khayr.2024}. The observed behavior implies that the
  dislocation density within the walls increases with strain, which provides a
  simple method to tune this important property.\@ (\textbf{D})-(\textbf{G})
  Diffuse streaks that originate from the long-range dislocation correlations
  within the walls.\@ (\textbf{D}) and (\textbf{F}) show neutron scattering
  measurements in two Brillouin zones, whereas (\textbf{E}) and (\textbf{G}) are
  the corresponding scattering intensities calculated for strain fields
  generated by periodic dislocation arrays~\cite{Hameed.2020,
  Khayr.2024}\label{plastic}.}
\end{figure}  

For a ceramic material, STO is strikingly ductile at ambient temperature, with
plastic deformation up to 10\% possible in compression. Moreover, the
deformation process leads to a remarkable self-organization of dislocations into
mesoscale structures, whose properties have recently been revealed using diffuse
scattering~\cite{Hameed.2020}. The most obvious effect in deformed single
crystals is an elongation of the usual Bragg peaks into arcs 
(Fig.~\ref{plastic}A), which are known as asterisms from early studies of 
plastically deformed metals~\cite{Gay.1951}. The presence of asterisms implies
that the sample breaks up into nearly unstrained tilted domains, with internal
strain concentrated around the domain boundaries (Fig.~\ref{plastic}B,C). The
local structure of the domain walls can be determined from weak diffuse streaks
that are observed away from the asterisms, which are fully consistent with
scattering from a periodic array of dislocations (Fig.~\ref{plastic}D-G). The
dislocations thus self-organize to form long-range periodic domain walls, with a
strain field that decays quickly away from the walls. This structural
information was essential to obtain a deeper understanding of the surprising
effects of plastic deformation on the electronic properties of STO, including
the appearance of quantum-critical ferroelectric fluctuations, a significant
boost of the superconducting $T_c$~\cite{Hameed.2020}, and emergent magnetism
and multiferroicity~\cite{Wang.2024}. 

The initial work on deformed STO has introduced the use of plastic deformation
to tune the properties of quantum materials, opening a new avenue in the field.
The perovskite KTaO$_3$ (KTO) was very recently shown to be ductile at ambient
temperature as well~\cite{Khayr.2024, Fang.2024}, with signatures of similar
structural and electronic features. Moreover, newly developed high-force
uniaxial strain cells have enabled pioneering x-ray diffuse scattering
experiments with \textit{in situ} applied stress, which have provided detailed
insights into the formation of asterisms in STO with increasing
strain~\cite{Khayr.2024}. Given the importance of uniaxial stress as a tuning
knob for the properties of quantum materials, such devices will be useful for
diffuse scattering measurements on a wide range of interesting material systems.

\subsection*{Stacking Faults in Layered Materials}

\begin{figure}[!b]
  \includegraphics[width=\columnwidth]{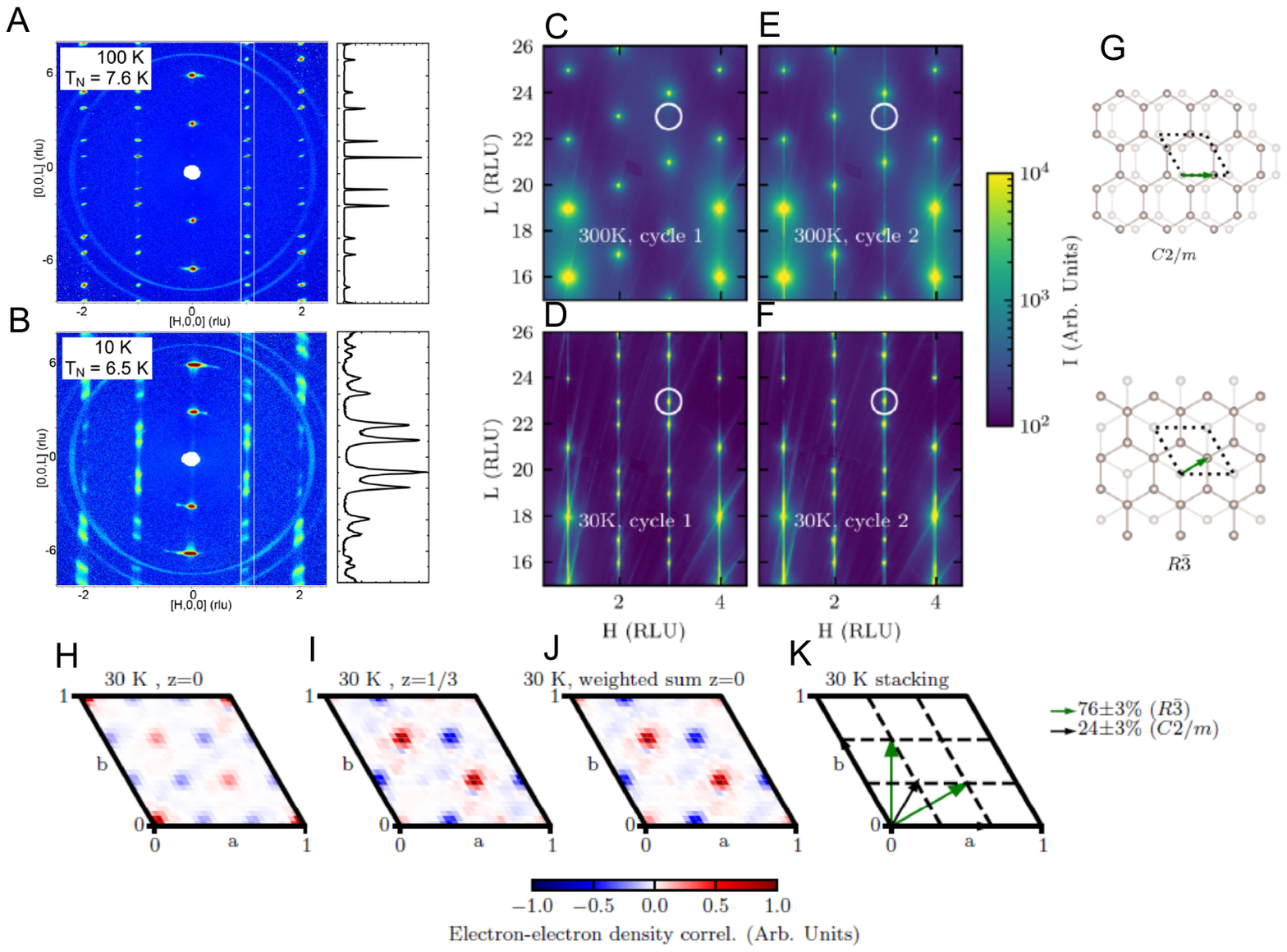} 
  \caption{Stacking faults in the Kitaev spin-liquid candidate material
  RuCl$_3$~\cite{Sears.2023a, Zhang.2024} (\textbf{A}, \textbf{B})
  Low-temperature neutron diffuse scattering (CORELLI) for two RuCl$_3$ crystals
  that show slightly different N\'eel temperatures, with clear differences in
  the stacking fault concentration~\cite{Zhang.2024}. The sample in (A) is
  nearly pristine, while the sample in (\textbf{B}) shows diffuse rods along the
  $L$ direction indicative of stacking faults.\@ (\textbf{C})-(\textbf{F}) X-ray
  diffuse scattering in a RuCl$_3$ crystal cycled through a structural
  transition that occurs between 100 and 200 K and rearranges the stacking
  sequence (as shown in (\textbf{G}))~\cite{Sears.2023a}. Stacking faults are
  absent in the as-grown sample, (\textbf{C}), while each cycle induces more
  defects both above and below the transition (\textbf{D})-(\textbf{F}).\@
  (\textbf{H})-(\textbf{J}) 3D-$\Delta$PDF generated from the x-ray scattering
  data. Cuts for two values of the out-of-plane coordinate $z$ are shown, $z =
  0$ (H) and $z = 1/3$ (I); the latter corresponds to the distance between two
  adjacent Ru-Cl planes. (J) Weighted sum of the $z = 0$ data shifted by two
  distinct vectors that correspond to the stacking sequences in the $R3$ and
  $C2/m$ structures. The agreement with (I) shows that the low-temperature
  structure is a mixture of the two sequences, with their concentrations denoted
  in (K).\label{stacking}}
\end{figure}

Many important materials exhibit irregularities in the stacking sequence of
their crystallographic planes. Such stacking faults are a specific type of
planar defect, and can play an important role in determining electronic
properties. Strongly anisotropic, layered materials with weak van der Waals
bonds between the layers are particularly susceptible, due to the low energies
needed to create stacking faults. Given their planar nature, these defects
typically lead to rod-like diffuse features, which can be used to determine both
their structure and concentration. Perhaps the most prominent recent scattering
work on stacking faults in quantum materials has been in the context of magnetic
systems such as RuCl$_3$~\cite{Sears.2023a, Zhang.2024} (Fig.~\ref{stacking}).
This material has been the subject of tremendous attention~\cite{Banerjee.2016,
Banerjee.2017}, since it is a candidate to host the elusive Kitaev quantum
spin-liquid state, which is expected to exhibit exotic excitations and holds
promise for quantum computation~\cite{Takagi.2019}. Yet due to residual
interactions between van der Waals bonded hexagonal RuCl$_3$ layers, the system
orders magnetically below about 10 K and does not show a spin-liquid ground
state, at least in the absence of an applied magnetic field. The long-range
magnetic order is exceedingly sensitive to the stacking
sequence~\cite{Banerjee.2016, Cao.2016}, which has motivated efforts to grow and
characterize crystals with ever smaller stacking fault concentrations. Neutron
and x-ray diffuse scattering have been indispensable in the efforts to quantify
the stacking fault density and uncover their interplay with other structural
features.

The stacking fault density in the highest-quality RuCl$_3$ crystals is
negligibly low at room temperature; however, the material undergoes a
first-order structural transition between 100 and 200 K, that involves a
rearrangement of the planar stacking~\cite{Sears.2023a,Zhang.2024}. Neutron
diffuse scattering measurements with \textit{CORELLI} have shown that the
sharpness of the structural transition strongly depends on the sample quality
and is correlated with the intensity of the diffuse rods associated with the
stacking faults~\cite{Zhang.2024} (Fig.~\ref{stacking}A,B). Electronic
properties such as the magnetic ordering transitions and spin thermal transport
are also sensitive to the stacking fault concentration. High-sensitivity x-ray
scattering experiments have demonstrated that even the best crystals acquire
stacking faults below the structural transition (Fig.~\ref{stacking}C-G), due to
an incomplete rearrangement of the layers and extremely low defect activation
energies~\cite{Sears.2023a}. Moreover, 3D-$\Delta$PDF analysis was successfully
used to provide insight into different stacking sequences and their relative
weights~\cite{Sears.2023a} (Fig.~\ref{stacking}H-K), which, to our knowledge, is
the first application of the method in studies of extended defects. This work
has provided clarity on the structural complexities of RuCl$_3$ that is
essential to understand the interplay between structure and magnetism. The
methodology is also relevant for a broad range of interesting layered materials,
from systems displaying magnetic or CDW order to exotic superconductors.

\section*{Outlook}

The previous sections have shown the novel insights into correlated electron
systems that existing single-crystal diffuse scattering capabilities can
provide. In this concluding section, we will briefly describe developments in
instrumentation, detectors, sample environments, and data analysis, which will
expand the scope of future scientific investigations.

While current x-ray diffuse scattering measurements over large volumes of
reciprocal space are limited to an approximate temperature range of 15~K to
700~K, many properties of interest in quantum and strongly correlated materials
require measurements to much lower temperatures, if possible into the mK range.
At the same time, extending the range to higher temperatures will also enable
experiments that probe the creation and evolution of short-range correlations at
temperatures approaching the melting point, as well as open new capabilities for
\textit{in situ} monitoring of high-temperature plastic deformation and
dislocation engineering.

The challenge for developing such capabilities will be to keep background
scattering to a minimum while still enabling measurements over large sample
rotation angles and, in the case of x-rays to avoid beam-heating of the sample
at low temperatures. Furthermore, the ability to simultaneously apply pressure
and magnetic or electric fields would enable detailed studies of correlated
disorder across multidimensional phase diagrams producing insights into quantum
phase transitions and other correlated electron phenomena. As discussed earlier,
uniaxial strain is another important parameter to tune or qualitatively modify
the properties of quantum materials in the elastic and plastic regimes. First
tests of a dedicated, specially designed uniaxial strain cell have been
performed at the Advanced Photon Source~\cite{Khayr.2024}, and a device that is
compatible with \textit{CORELLI} is under development as well.

Low- and high-temperature sample environments and magnetic fields are in
principle all available for diffuse neutron scattering measurements, although
the required sample sizes may impact crystal quality. Even with larger samples,
measurement times are still an order of magnitude longer than for x-ray
experiments, limiting investigations to a few points in the phase diagram. This
will require new instrumentation at future facilities, such as the
time-of-flight Laue instrument \textit{PIONEER}~\cite{Liu.2022} proposed to be
built at the Second Target Station of the Spallation Neutron Source, optimized
for small samples in the range of 0.1 to 1~mm and large detector coverage of a
solid angle of about 4~sr.

Developments in x-ray detector technologies promise to enable higher frame rates
with higher dynamic range. Both of these are important to reduce artefacts due
to strong Bragg peaks that can only be removed by elaborate
procedures~\cite{Welberry.2022, Krogstad.2019tc, Koch.2021}, which can still
leave some spurious signals in the transformed data. Recent and planned updates
to synchrotron facilities resulting in smaller beam sizes and higher brilliance,
combined with higher detector frame rates, open the possibility for time- and
spatially-resolved measurements. This could enable, \textit{e.g.},
scanning-probe diffuse scattering measurements of inhomogeneities at the micron
size. Stroboscopic measurements could further probe the presence and response of
correlated disorder in electric field-driven states.

Finally, optimizing the extraction of all the information contained in complete
data sets requires further developments of tools for automated feature detection
and physically interpretable models of correlated disorder embedded in a
long-range ordered crystalline lattice. Traditional methods are based on
parametrizing the diffuse scattering with Warren-Cowley
parameters~\cite{Welberry.2022}, performing Monte Carlo (MC) simulations of
effective Hamiltonians~\cite{Welberry.2008, Weber.2002}, or using reverse Monte
Carlo (RMC) simulations to generate real-space structures~\cite{Nield.1995,
Welberry.1998}. The MC method has the advantage that it generally only involves
relatively few interatomic interaction parameters and provides direct physical
insights. However, it requires a specific model to be tailored for each system
with parameters based on known physical and chemical principles. In the RMC
method on the other hand, the adjustable parameters are the positions of all the
atoms in a box, which does not provide direct physical insight without further
statistical analysis to obtain, \textit{e.g.}, conditional probabilities.
3D-$\Delta$PDF has recently been incorporated into RMC approaches either to
build a starting model for the simulations, \textit{e.g.}, in the program
\textit{YELL}~\cite{Simonov.2014}, or to help in validating and interpreting RMC
results~\cite{Morgan.2021}.

Future developments could further use machine learning to derive the structure
of short-range order directly from measured 3D-$\Delta$PDF maps, as has already
been performed for long-range ordered structures~\cite{Pan.2023}. In addition,
symmetry-mode analysis methods can reduce the complexity of describing
distortions derived from higher symmetry phases~\cite{Hamilton.2023}, and
mean-field approaches have been implemented for efficient fitting of
single-crystal diffuse scattering data~\cite{Schmidt.2022}. Ultimately, large
supervised ML models incorporating such various approaches in real-, Patterson,
and reciprocal space to obtain physical models could lead to a much more user
friendly analysis of single crystal diffuse scattering, on a par with standard
PDF analysis. This would almost certainly lead to the more widespread adoption
of the techniques described here in future investigations of the role of
inhomogeneity in the properties of correlated electron materials.

\subsection*{Acknowledgements}

We acknowledge helpful contributions from Eun-Ah Kim, Chris Leighton, and
Krishnanand Mallayya. The work at Argonne was supported by the U.S. Department
of Energy, Office of Science, Basic Energy Sciences, Materials Sciences and
Engineering Division. The work at the University of Minnesota was supported by
the U.S. Department of Energy through the University of Minnesota Center for
Quantum Materials, under grant number DE-SC-0016371. The work at the University
of Zagreb was supported by the Croatian Science Foundation under grant number
UIP-2020-02-9494, and the Croatian Ministry of Science and Education.

\bibliography{review}

\begin{thebibliography}{100}

\bibitem{Aeppli.1997ju}
G.~Aeppli, P.~Chandra, {Seeking a simple complex system}.
\newblock {\it Science\/} {\bf 275}, 177--178 (1997).

\bibitem{Dagotto.2005ip}
E.~Dagotto, {Complexity in strongly correlated electronic systems}.
\newblock {\it Science\/} {\bf 309}, 257--262 (2005).

\bibitem{Keen.2015bq}
D.~A. Keen, A.~L. Goodwin, {The crystallography of correlated disorder}.
\newblock {\it Nature\/} {\bf 521}, 303--309 (2015).

\bibitem{Mazza.2024}
A.~R. Mazza, J.-Q. Yan, S.~Middey, J.~S. Gardner, A.-H. Chen, M.~Brahlek, T.~Z. Ward, {Embracing disorder in quantum materials design}.
\newblock {\it Appl. Phys. Lett.\/} {\bf 124}, 230501 (2024).

\bibitem{Zhou.2017dk}
Y.~Zhou, K.~Kanoda, T.-K. Ng, {Quantum spin liquid states}.
\newblock {\it Rev. Mod. Phys.\/} {\bf 89}, 025003 (2017).

\bibitem{Broholm.2020cc}
C.~Broholm, R.~J. Cava, S.~A. Kivelson, D.~G. Nocera, M.~R. Norman, T.~Senthil, Quantum spin liquids.
\newblock {\it Science\/} {\bf 367}, eaay0668 (2020).

\bibitem{Tokura.2006ff}
Y.~Tokura, {Critical features of colossal magnetoresistive manganites}.
\newblock {\it Rep. Prog. Phys.\/} {\bf 69}, 797 (2006).

\bibitem{Lee.2006de}
P.~A. Lee, N.~Nagaosa, X.-G. Wen, {Doping a Mott insulator: Physics of high-temperature superconductivity}.
\newblock {\it Rev. Mod. Phys.\/} {\bf 78}, 17 (2006).

\bibitem{Canfield.2010}
P.~C. Canfield, S.~L. Bud'ko, {FeAs-Based Superconductivity: A Case Study of the Effects of Transition Metal Doping on BaFe$_2$As$_2$}.
\newblock {\it Annu. Rev. Condens. Matter Phys.\/} {\bf 1}, 27--50 (2010).

\bibitem{Vojta.2003}
M.~Vojta, {Quantum phase transitions}.
\newblock {\it Rep. Prog. Phys.\/} {\bf 66}, 2069--2110 (2003).

\bibitem{vanderMarel.2003cf}
D.~v.~d. Marel, H.~Molegraaf, J.~Zaanen, Z.~Nussinov, F.~Carbone, A.~Damascelli, H.~Eisaki, M.~Greven, P.~H. Kes, M.~Li, {Quantum critical behaviour in a high-T-c superconductor}.
\newblock {\it Nature\/} {\bf 425}, 271--274 (2003).

\bibitem{Shibauchi.2014bw}
T.~Shibauchi, A.~Carrington, Y.~Matsuda, {A Quantum Critical Point Lying Beneath the Superconducting Dome in Iron Pnictides}.
\newblock {\it Annu. Rev. Condens. Matter Phys.\/} {\bf 5}, 113--135 (2014).

\bibitem{Billinge.2004uu}
S.~J.~L. Billinge, {The atomic pair distribution function: past and present}.
\newblock {\it Z. Krist.\/} {\bf 219}, 117--121 (2004).

\bibitem{Keen.2020}
D.~A. Keen, {Total scattering and the pair distribution function in crystallography}.
\newblock {\it Crystallogr. Rev.\/} {\bf 26}, 143--201 (2020).

\bibitem{Farrow.2007}
C.~L. Farrow, P.~Juhas, J.~W. Liu, D.~Bryndin, E.~S. Božin, J.~Bloch, T.~Proffen, S.~J.~L. Billinge, \textit{PDFfit2} and \textit{PDFgui}: computer programs for studying nanostructure in crystals.
\newblock {\it J. Phys. Cond. Matt.\/} {\bf 19}, 335219 (2007).

\bibitem{Welberry.2022}
R.~Welberry, {\it {Diffuse X-ray Scattering and Models of Disorder}\/} (Oxford University Press, Oxford, UK, 2022).

\bibitem{Nield.2001wg}
V.~M. Nield, D.~A. Keen, {\it {Diffuse neutron scattering from crystalline materials}\/} (Oxford University Press, Oxford, UK, 2001).

\bibitem{Krogstad.2019tc}
M.~J. Krogstad, S.~Rosenkranz, J.~M. Wozniak, G.~Jennings, J.~P.~C. Ruff, J.~T. Vaughey, R.~Osborn, {Reciprocal Space Imaging of Ionic Correlations in Intercalation Compounds}.
\newblock {\it Nat. Mater.\/} {\bf 19}, 63--68 (2020).

\bibitem{Ye.2018bc}
F.~Ye, Y.~Liu, R.~Whitfield, R.~Osborn, S.~Rosenkranz, {Implementation of cross correlation for energy discrimination on the time-of-flight spectrometer \textit{CORELLI}}.
\newblock {\it J. Appl. Cryst.\/} {\bf 51}, 315--322 (2018).

\bibitem{Proffen.1997fc}
T.~Proffen, R.~B. Neder, {\textit{DISCUS}: a program for diffuse scattering and defect-structure simulation}.
\newblock {\it J. Appl. Cryst.\/} {\bf 30}, 171--175 (1997).

\bibitem{Weber.2012en}
T.~Weber, A.~Simonov, {The three-dimensional pair distribution function analysis of disordered single crystals: basic concepts}.
\newblock {\it Z. Krist.\/} {\bf 227}, 238--247 (2012).

\bibitem{Osborn.2023}
R.~Osborn, {Mapping structural correlations in real space}.
\newblock {\it Acta Cryst. B\/} {\bf 79}, 99--100 (2023).

\bibitem{Upreti.2022}
P.~Upreti, M.~Krogstad, C.~Haley, M.~Anitescu, V.~Rao, L.~Poudel, O.~Chmaissem, S.~Rosenkranz, R.~Osborn, {Order-Disorder Transitions in (Ca$_x$Sr$_{1-x}$)$_3$Rh$_4$Sn$_{13}$}.
\newblock {\it Phys. Rev. Lett.\/} {\bf 128}, 095701 (2022).

\bibitem{Venderley.2022}
J.~Venderley, M.~Matty, K.~Mallaya, M.~Krogstad, J.~Ruff, G.~Pleiss, V.~Kishore, D.~Mandrus, D.~Phelan, A.~G. Wilson, K.~Weinberger, P.~Upreti, M.~R. Norman, S.~Rosenkranz, R.~Osborn, E.-A. Kim, {Harnessing interpretable and unsupervised machine learning to address big data from modern X-ray diffraction}.
\newblock {\it Proc. Natl. Acad. Sci. U.S.A.\/} {\bf 119}, e2109665119 (2022).

\bibitem{Kautzsch.2023}
L.~Kautzsch, B.~R. Ortiz, K.~Mallayya, J.~Plumb, G.~Pokharel, J.~P.~C. Ruff, Z.~Islam, E.-A. Kim, R.~Seshadri, S.~D. Wilson, {Structural evolution of the kagome superconductors AV$_3$Sb$_5$ (A = K, Rb, and Cs) through charge density wave order}.
\newblock {\it Phys. Rev. Mater.\/} {\bf 7}, 024806 (2023).

\bibitem{Pokharel.2023}
G.~Pokharel, B.~R. Ortiz, L.~Kautzsch, S.~J.~A. Gomez, K.~Mallayya, G.~Wu, E.-A. Kim, J.~P.~C. Ruff, S.~Sarker, S.~D. Wilson, {Frustrated charge order and cooperative distortions in ScV$_6$Sn$_6$}.
\newblock {\it Phys. Rev. Mater.\/} {\bf 7}, 104201 (2023).

\bibitem{Mallayya.2024}
K.~Mallayya, J.~Straquadine, M.~J. Krogstad, M.~D. Bachmann, A.~G. Singh, R.~Osborn, S.~Rosenkranz, I.~R. Fisher, E.-A. Kim, {Bragg glass signatures in Pd$_x$ErTe$_3$ with X-ray diffraction temperature clustering}.
\newblock {\it Nat. Phys.\/} pp. 1--8 (2024).

\bibitem{Patterson.1934}
A.~L. Patterson, {A Fourier Series Method for the Determination of the Components of Interatomic Distances in Crystals}.
\newblock {\it Phys. Rev.\/} {\bf 46}, 372--376 (1934).

\bibitem{Sangiorgio.2018gj}
B.~Sangiorgio, E.~S. Bo{\v z}in, C.~D. Malliakas, M.~Fechner, A.~Simonov, M.~G. Kanatzidis, S.~J.~L. Billinge, N.~A. Spaldin, T.~Weber, {Correlated local dipoles in PbTe}.
\newblock {\it Phys. Rev. Mater.\/} {\bf 2}, 085402 (2018).

\bibitem{Holm.2020}
K.~A.~U. Holm, N.~Roth, C.~M. Zeuthen, K.~Tolborg, A.~A. Feidenhans'l, B.~B. Iversen, Temperature dependence of dynamic dipole formation in pbte.
\newblock {\it Phys. Rev. B\/} {\bf 102}, 024112 (2020).

\bibitem{Stockler.2022}
K.~A.~H. Støckler, N.~Roth, T.~B.~E. Grønbech, B.~B. Iversen, {Epitaxial intergrowths and local oxide relaxations in natural bixbyite Fe$_{2-x}$Mn$_x$O$_3$}.
\newblock {\it IUCrJ\/} {\bf 9}, 523--532 (2022).

\bibitem{Roth.2019}
N.~Roth, F.~Ye, A.~F. May, B.~C. Chakoumakos, B.~B. Iversen, Magnetic correlations and structure in bixbyite across the spin-glass transition.
\newblock {\it Phys. Rev. B\/} {\bf 100}, 144404 (2019).

\bibitem{Davenport.2019}
M.~A. Davenport, M.~J. Krogstad, L.~M. Whitt, S.~Rosenkranz, R.~Osborn, J.~M. Allred, {Two-dimensional ordering phase brought on by the destabilization of the VO$_2$ rutile structure in V$_{0.81}$Mo$_{0.19}$O$_2$}.
\newblock {\it Acta Cryst. A\/} {\bf 75}, a236 (2019).

\bibitem{Imada.1998ti}
M.~Imada, A.~Fujimori, Y.~Tokura, {Metal-insulator transitions}.
\newblock {\it Rev. Mod. Phys.\/} {\bf 70}, 1039--1263 (1998).

\bibitem{Georgescu.2022}
A.~B. Georgescu, A.~J. Millis, {Quantifying the role of the lattice in metal--insulator phase transitions}.
\newblock {\it Commun. Phys.\/} {\bf 5}, 135 (2022).

\bibitem{Yang.2016}
M.~Yang, Y.~Yang, B.~Hong, L.~Wang, K.~Hu, Y.~Dong, H.~Xu, H.~Huang, J.~Zhao, H.~Chen, L.~Song, H.~Ju, J.~Zhu, J.~Bao, X.~Li, Y.~Gu, T.~Yang, X.~Gao, Z.~Luo, C.~Gao, {Suppression of Structural Phase Transition in VO$_2$ by Epitaxial Strain in Vicinity of Metal-insulator Transition}.
\newblock {\it Sci. Rep.\/} {\bf 6}, 23119 (2016).

\bibitem{Ji.2021}
Y.~Ji, L.~Cheng, N.~Li, Y.~Yuan, W.~Liang, H.~Yang, {Decoupling between metal--insulator transition and structural phase transition in an interface-engineered VO$_2$}.
\newblock {\it J. Phys. Cond. Matt.\/} {\bf 33}, 105603 (2021).

\bibitem{Goodenough.1971dz}
J.~B. Goodenough, {The two components of the crystallographic transition in VO$_2$}.
\newblock {\it J. Solid State Chem.\/} {\bf 3}, 490--500 (1971).

\bibitem{Pouget.1974fv}
J.~P. Pouget, H.~Launois, T.~M. Rice, P.~Dernier, A.~Gossard, G.~Villeneuve, P.~Hagenmuller, {Dimerization of a linear Heisenberg chain in the insulating phases of V$_{1-x}$Cr$_x$O$_2$}.
\newblock {\it Phys. Rev. B\/} {\bf 10}, 1801--1815 (1974).

\bibitem{Rice.1994fr}
T.~M. Rice, H.~Launois, J.~P. Pouget, {Comment on "VO$_2$: Peierls or Mott-Hubbard? A View from Band Theory"}.
\newblock {\it Phys. Rev. Lett.\/} {\bf 73}, 3042--3042 (1994).

\bibitem{Holman.2009ft}
K.~L. Holman, T.~M. McQueen, A.~J. Williams, T.~Klimczuk, P.~W. Stephens, H.~W. Zandbergen, Q.~Xu, F.~Ronning, R.~J. Cava, {Insulator to correlated metal transition in V$_{1-x}$Mo$_x$O$_2$}.
\newblock {\it Phys. Rev. B\/} {\bf 79}, 245114 (2009).

\bibitem{Chhetri.2022}
T.~B.~R. Chhetri, T.~C. Douglas, M.~A. Davenport, S.~Rosenkranz, R.~Osborn, M.~J. Krogstad, J.~M. Allred, {Geometric Frustration Suppresses Long-Range Structural Distortions in Nb$_x$V$_{1-x}$O$_2$}.
\newblock {\it J. Phys. Chem. C\/} {\bf 126}, 2049--2061 (2022).

\bibitem{Bussmann-Holder.2006}
A.~Bussmann-Holder, N.~Dalal, {\it Structure and Bonding\/} (Springer, Berlin, Heidelberg, 2006), vol. 124, pp. 1--21.

\bibitem{Klintberg.2012dt}
L.~E. Klintberg, S.~K. Goh, P.~L. Alireza, P.~J. Saines, D.~A. Tompsett, P.~W. Logg, J.~Yang, B.~Chen, K.~Yoshimura, F.~M. Grosche, {Pressure- and Composition-Induced Structural Quantum Phase Transition in the Cubic Superconductor Sr$_3$Ir$_4$Sn$_{13}$}.
\newblock {\it Phys. Rev. Lett.\/} {\bf 109}, 237008 (2012).

\bibitem{Ban.2017}
W.~J. Ban, H.~P. Wang, C.~W. Tseng, C.~N. Kuo, C.~S. Lue, N.~L. Wang, {Optical spectroscopy study of charge density wave order in Sr$_3$Rh$_4$Sn$_{13}$ and (Sr$_{0.5}$Ca$_{0.5}$)$_3$Rh$_4$Sn$_{13}$}.
\newblock {\it Sci. China Phys. Mech.\/} {\bf 60}, 047011 (2017).

\bibitem{Chatterjee.2015cw}
U.~Chatterjee, J.~Zhao, M.~Iavarone, R.~D. Capua, J.-P. Castellan, G.~Karapetrov, C.~D. Malliakas, M.~G. Kanatzidis, H.~Claus, J.~P.~C. Ruff, F.~Weber, J.~v. Wezel, J.-C. Campuzano, R.~Osborn, M.~Randeria, N.~Trivedi, M.~R. Norman, S.~Rosenkranz, {Emergence of coherence in the charge-density wave state of 2$H$-NbSe$_2$}.
\newblock {\it Nat. Commun.\/} {\bf 6}, 6313 (2015).

\bibitem{Onodera.2004}
Y.~Onodera, {Dynamical Response of Ferroelectrics in Terms of a Classical Anharmonic-Oscillator Model}.
\newblock {\it J. Phys. Soc. Japan\/} {\bf 73}, 1216--1221 (2004).

\bibitem{Fu.2015cl}
L.~Fu, {Parity-Breaking Phases of Spin-Orbit-Coupled Metals with Gyrotropic, Ferroelectric, and Multipolar Orders}.
\newblock {\it Phys. Rev. Lett.\/} {\bf 115}, 026401 (2015).

\bibitem{Cao.2018be}
G.~Cao, P.~Schlottmann, {The challenge of spin-orbit-tuned ground states in iridates: a key issues review}.
\newblock {\it Rep. Prog. Phys.\/} {\bf 81}, 042502 (2018).

\bibitem{Harter.2017dr}
J.~W. Harter, Z.~Y. Zhao, J.-Q. Yan, D.~G. Mandrus, D.~Hsieh, {A parity-breaking electronic nematic phase transition in the spin-orbit coupled metal Cd$_2$Re$_2$O$_7$}.
\newblock {\it Science\/} {\bf 356}, 295--299 (2017).

\bibitem{Norman.2020hn}
M.~R. Norman, {Crystal structure of the inversion-breaking metal Cd$_2$Re$_2$O$_7$}.
\newblock {\it Phys. Rev. B\/} {\bf 101}, 045117 (2020).

\bibitem{Yamaura.2002fk}
J.-I. Yamaura, Z.~Hiroi, {Low Temperature Symmetry of Pyrochlore Oxide Cd$_2$Re$_2$O$_7$}.
\newblock {\it J. Phys. Soc. Japan\/} {\bf 71}, 2598--2600 (2002).

\bibitem{Castellan.2002kz}
J.-P. Castellan, B.~D. Gaulin, J.~v. Duijn, M.~J. Lewis, M.~D. Lumsden, R.~Jin, J.~He, S.~E. Nagler, D.~Mandrus, {Structural ordering and symmetry breaking in Cd$_2$Re$_2$O$_7$}.
\newblock {\it Phys. Rev. B\/} {\bf 66}, 134528 (2002).

\bibitem{Weller.2004gf}
M.~T. Weller, R.~W. Hughes, J.~Rooke, C.~S. Knee, J.~Reading, {The pyrochlore family -- a potential panacea for the frustrated perovskite chemist}.
\newblock {\it Dalton Trans.\/} pp. 3032--3041 (2004).

\bibitem{Imry.1975bi}
Y.~Imry, S.-k. Ma, {Random-Field Instability of the Ordered State of Continuous Symmetry}.
\newblock {\it Phys. Rev. Lett.\/} {\bf 35}, 1399--1401 (1975).

\bibitem{Fukuyama.1978}
H.~Fukuyama, P.~A. Lee, {Dynamics of the charge-density wave. I. Impurity pinning in a single chain}.
\newblock {\it Phys. Rev. B\/} {\bf 17}, 535--541 (1978).

\bibitem{Nattermann.1989}
T.~Nattermann, {Scaling approach to pinning: Charge density waves and giant flux creep in superconductors}.
\newblock {\it Phys. Rev. Lett.\/} {\bf 64}, 2454--2457 (1990).

\bibitem{Giamarchi.1994}
T.~Giamarchi, P.~L. Doussal, {Elastic theory of pinned flux lattices}.
\newblock {\it Phys. Rev. Lett.\/} {\bf 72}, 1530--1533 (1994).

\bibitem{Giamarchi.1995}
T.~Giamarchi, P.~L. Doussal, {Elastic theory of flux lattices in the presence of weak disorder}.
\newblock {\it Phys. Rev. B\/} {\bf 52}, 1242--1270 (1995).

\bibitem{Okamoto.2015vo}
J.-i. Okamoto, C.~J. Arguello, E.~P. Rosenthal, A.~N. Pasupathy, A.~J. Millis, {Experimental Evidence for a Bragg Glass Density Wave Phase in a Transition-Metal Dichalcogenide}.
\newblock {\it Phys. Rev. Lett.\/} {\bf 114}, 026802 (2015).

\bibitem{Fang.2019ha}
A.~Fang, J.~A.~W. Straquadine, I.~R. Fisher, S.~A. Kivelson, A.~Kapitulnik, {Disorder-induced suppression of charge density wave order: STM study of Pd-intercalated ErTe$_3$}.
\newblock {\it Phys. Rev. B\/} {\bf 100}, 235446 (2019).

\bibitem{Straquadine.2019ju}
J.~A.~W. Straquadine, F.~Weber, S.~Rosenkranz, A.~H. Said, I.~R. Fisher, {Suppression of charge density wave order by disorder in Pd-intercalated ErTe$_3$}.
\newblock {\it Phys. Rev. B\/} {\bf 99}, 235138 (2019).

\bibitem{Keimer.2015}
B.~Keimer, S.~A. Kivelson, M.~R. Norman, S.~Uchida, J.~Zaanen, {From quantum matter to high-temperature superconductivity in copper oxides}.
\newblock {\it Nature\/} {\bf 518}, 179--186 (2015).

\bibitem{Sleight.2015}
A.~W. Sleight, {Bismuthates: BaBiO$_3$ and related superconducting phases}.
\newblock {\it Physica C\/} {\bf 514}, 152--165 (2015).

\bibitem{Tranquada.1995}
J.~M. Tranquada, B.~J. Sternlieb, J.~D. Axe, Y.~Nakamura, S.~Uchida, {Evidence for stripe correlations of spins and holes in copper oxide superconductors}.
\newblock {\it Nature\/} {\bf 375}, 561--563 (1995).

\bibitem{Ghiringhelli.2012}
G.~Ghiringhelli, M.~Le~Tacon, M.~Minola, C.~Mazzoli, N.~B. Brookes, G.~M. De~Luca, A.~Frano, D.~G. Hawthorn, F.~He, T.~Loew, M.~M. Sala, D.~C. Peets, M.~Salluzzo, E.~Schierle, R.~Sutarto, G.~A. Sawatzky, E.~Weschke, B.~Keimer, L.~Braicovich, {Long-range incommensurate charge fluctuations in (Y,Nd)Ba$_2$Cu$_3$O$_{6+\delta}$}.
\newblock {\it Science\/} {\bf 337}, 821--825 (2012).

\bibitem{Chang.2012}
J.~Chang, E.~Blackburn, A.~T. Holmes, N.~B. Christensen, J.~Larsen, J.~Mesot, R.~Liang, D.~A. Bonn, W.~N. Hardy, A.~Watenphul, M.~von Zimmermann, E.~M. Forgan, S.~M. Hayden, {Direct observation of competition between superconductivity and charge density wave order in YBa$_2$Cu$_3$O$_{6.67}$}.
\newblock {\it Nat. Phys.\/} {\bf 8}, 871--876 (2012).

\bibitem{Comin.2014}
R.~Comin, A.~Frano, M.~M. Yee, Y.~Yoshida, H.~Eisaki, E.~Schierle, E.~Weschke, R.~Sutarto, F.~He, A.~Soumyanarayanan, Y.~He, M.~Le~Tacon, I.~S. Elfimov, J.~E. Hoffman, G.~A. Sawatzky, B.~Keimer, A.~Damascelli, {Charge order driven by Fermi-arc instability in Bi$_2$Sr$_{2-x}$La$_x$CuO$_{6+\delta}$}.
\newblock {\it Science\/} {\bf 343}, 390--392 (2014).

\bibitem{daSilvaNeto.2015}
E.~H. d.~S. Neto, R.~Comin, F.~He, R.~Sutarto, Y.~Jiang, R.~L. Greene, G.~A. Sawatzky, A.~Damascelli, {Charge ordering in the electron-doped superconductor Nd$_{2-x}$Ce$_x$CuO$_{4}$}.
\newblock {\it Science\/} {\bf 347}, 282--285 (2015).

\bibitem{Arpaia.2019}
R.~Arpaia, S.~Caprara, R.~Fumagalli, G.~De~Vecchi, Y.~Y. Peng, E.~Andersson, D.~Betto, G.~M. De~Luca, N.~B. Brookes, F.~Lombardi, M.~Salluzzo, L.~Braicovich, C.~D. Castro, M.~Grilli, G.~Ghiringhelli, {Dynamical charge density fluctuations pervading the phase diagram of a Cu-based high-$T_c$ superconductor}.
\newblock {\it Science\/} {\bf 365}, 906--910 (2019).

\bibitem{Yu.2020}
B.~Yu, W.~Tabis, I.~Bialo, F.~Yakhou, N.~B. Brookes, Z.~A. Anderson, Y.~Tang, G.~Yu, M.~Greven, {Unusual dynamic charge correlations in simple-tetragonal {HgBa$_2$CuO$_{4+\delta}$}}.
\newblock {\it Phys. Rev. X\/} {\bf 10}, 021059 (2020).

\bibitem{Griffitt.2023}
S.~Griffitt, M.~Spaić, J.~Joe, Z.~W. Anderson, D.~Zhai, M.~J. Krogstad, R.~Osborn, D.~Pelc, M.~Greven, {Local inversion-symmetry breaking in a bismuthate high-T$_c$ superconductor}.
\newblock {\it Nat. Commun.\/} {\bf 14}, 845 (2023).

\bibitem{Anderson.2024}
Z.~A. Anderson, M.~Spai\'{c}, N.~Biniskos, L.~Thompson, B.~Yu, J.~Zwettler, Y.~Liu, F.~Ye, G.~E. Granroth, M.~J. Krogstad, R.~Osborn, D.~Pelc, M.~Greven, {Bulk nanoscale structural correlations in a model cuprate superconductor}.
\newblock {\it arXiv:2405.10411\/}  (2024).

\bibitem{Cava.1988}
R.~J. Cava, B.~Batlogg, J.~J. Krajewski, R.~Farrow, L.~W. Rupp, A.~E. White, K.~Short, W.~F. Peck, T.~Kometani, {Superconductivity near 30 K without copper: the Ba$_{0.6}$K$_{0.4}$BiO$_3$ perovskite}.
\newblock {\it Nature\/} {\bf 332}, 814--816 (1988).

\bibitem{Pei.1990}
S.~Pei, J.~D. Jorgensen, B.~Dabrowski, D.~G. Hinks, D.~R. Richards, A.~W. Mitchell, J.~M. Newsam, S.~K. Sinha, D.~Vaknin, A.~J. Jacobson, {Structural phase diagram of the Ba$_{1-x}$K$_x$BiO$_3$ system}.
\newblock {\it Phys. Rev. B\/} {\bf 41}, 4126--4141 (1990).

\bibitem{Jurczek.1986}
E.~Jurczek, T.~M. Rice, {A Charge-Density-Wave Instability in BaBi$_{1-x}$Pb$_x$O$_3$ Caused by Strong Electron-Phonon Coupling}.
\newblock {\it EPL\/} {\bf 1}, 225--231 (1986).

\bibitem{Jiang.2021}
M.~Jiang, G.~A. Sawatzky, M.~Berciu, S.~Johnston, {Polaron and bipolaron tendencies in a semiclassical model for hole-doped bismuthates}.
\newblock {\it Phys. Rev. B\/} {\bf 103}, 115129 (2021).

\bibitem{Yin.2013}
Z.~P. Yin, A.~Kutepov, G.~Kotliar, {Correlation-Enhanced Electron-Phonon Coupling: Applications of GW and Screened Hybrid Functional to Bismuthates, Chloronitrides, and Other High-T$_c$ Superconductors}.
\newblock {\it Phys. Rev. X\/} {\bf 3}, 021011 (2013).

\bibitem{Whitfield.2014}
R.~E. Whitfield, T.~R. Welberry, M.~Pa{\'{s}}ciak, D.~J. Goossens, {Use of bond-valence sums in modelling the diffuse scattering from PZN (PbZn$_{1/3}$Nb$_{2/3}$O$_3$)}.
\newblock {\it Acta Cryst. A\/} {\bf 70}, 626--635 (2014).

\bibitem{Smidman.2017}
M.~Smidman, M.~B. Salamon, H.~Q. Yuan, D.~F. Agterberg, {Superconductivity and spin–orbit coupling in non-centrosymmetric materials: a review}.
\newblock {\it Rep. Prog. Phys.\/} {\bf 80}, 036501 (2017).

\bibitem{Kim.2022}
M.~Kim, G.~M. McNally, H.-H. Kim, M.~Oudah, A.~S. Gibbs, P.~Manuel, R.~J. Green, R.~Sutarto, T.~Takayama, A.~Yaresko, U.~Wedig, M.~Isobe, R.~K. Kremer, D.~A. Bonn, B.~Keimer, H.~Takagi, {Superconductivity in (Ba,K)SbO$_3$}.
\newblock {\it Nat. Mater.\/} {\bf 21}, 627--633 (2022).

\bibitem{Sears.2023}
J.~Sears, Y.~Shen, M.~J. Krogstad, H.~Miao, E.~S. Bo{\v z}in, I.~K. Robinson, G.~D. Gu, R.~Osborn, S.~Rosenkranz, J.~M. Tranquada, M.~P.~M. Dean, {Structure of charge density waves in La$_{1.875}$Ba$_{0.125}$CuO$_4$}.
\newblock {\it Phys. Rev. B\/} {\bf 107}, 115125 (2023).

\bibitem{Pelc.2022}
D.~Pelc, R.~J. Spieker, Z.~W. Anderson, M.~J. Krogstad, N.~Biniskos, N.~G. Bielinski, B.~Yu, T.~Sasagawa, L.~Chauviere, P.~Dosanjh, R.~Liang, D.~A. Bonn, A.~Damascelli, S.~Chi, Y.~Liu, R.~Osborn, M.~Greven, {Unconventional short-range structural fluctuations in cuprate superconductors}.
\newblock {\it Sci. Rep.\/} {\bf 12}, 20483 (2022).

\bibitem{Egami.1994}
T.~Egami, S.~J.~L. Billinge, {Lattice effects in high-temperature superconductors}.
\newblock {\it Prog. Mater. Sci.\/} {\bf 38}, 359--424 (1994).

\bibitem{Pan.2001}
S.~H. Pan, J.~P. O'Neal, R.~L. Badzey, C.~Chamon, H.~Ding, J.~R. Engelbrecht, Z.~Wang, H.~Eisaki, S.~Uchida, A.~K. Gupta, K.-W. Ng, E.~W. Hudson, K.~M. Lang, J.~C. Davis, {Microscopic electronic inhomogeneity in the high-$T_c$ superconductor Bi$_2$Sr$_2$CaCu$_2$O$_{8+x}$}.
\newblock {\it Nature\/} {\bf 413}, 282--285 (2001).

\bibitem{Fischer.2007}
{\O}.~Fischer, M.~Kugler, I.~Maggio-Aprile, C.~Berthod, C.~Renner, {Scanning tunneling spectroscopy of high temperature superconductors}.
\newblock {\it Rev. Mod. Phys.\/} {\bf 79}, 353--419 (2007).

\bibitem{Du.2020}
Z.~Du, H.~Li, S.~H. Joo, E.~P. Donoway, J.~Lee, J.~C.~S. Davis, G.~Gu, P.~D. Johnson, K.~Fujita, { Imaging the energy gap modulations of the cuprate pair-density-wave state}.
\newblock {\it Nature\/} {\bf 580}, 65--70 (2020).

\bibitem{Singer.2002}
P.~M. Singer, A.~W. Hunt, T.~Imai, {$^{63}$Cu NQR evidence for spatial variation of hole concentration in L${\mathrm{a}}_{2-x}$S${\mathrm{r}}_{x}$Cu${\mathrm{O}}_{4}$}.
\newblock {\it Phys. Rev. Lett.\/} {\bf 88}, 047602 (2002).

\bibitem{Bobroff.2002}
J.~Bobroff, H.~Alloul, S.~Ouazi, P.~Mendels, A.~Mahajan, N.~Blanchard, G.~Collin, V.~Guillen, J.-F. Marucco, { Absence of static phase separation in the high $T_c$ cuprate YBa$_2$Cu$_3$O$_{6+x}$}.
\newblock {\it Phys. Rev. Lett\/} {\bf 89}, 157002 (2002).

\bibitem{Frachet.2020}
M.~Frachet, I.~Vinograd, R.~Zhou, S.~Benhabib, S.~Wu, H.~Mayaffre, S.~Kr{\"a}mer, S.~K. Ramakrishna, A.~P. Reyes, J.~Debray, T.~Kurosawa, N.~Momono, M.~Oda, S.~Komiya, S.~Ono, M.~Horio, J.~Chang, C.~Proust, L.~D, M.-H. Julien, {Hidden magnetism at the pseudogap critical point of a cuprate superconductor}.
\newblock {\it Nat. Phys.\/} {\bf 16}, 1064--1068 (2020).

\bibitem{Bianconi.1996}
A.~Bianconi, N.~L. Saini, A.~Lanzara, M.~Missori, T.~Rossetti, H.~Oyanagi, H.~Yamaguchi, K.~Oka, T.~Ito, {Determination of the local lattice distortions in the Cu${\mathrm{O}}_{2}$ plane of L${\mathrm{a}}_{1.85}$S${\mathrm{r}}_{0.15}$Cu${\mathrm{O}}_{4}$}.
\newblock {\it Phys. Rev. Lett.\/} {\bf 76}, 3412--3415 (1996).

\bibitem{Fratini.2010}
M.~Fratini, N.~Poccia, A.~Ricci, G.~Campi, M.~Burghammer, G.~Aeppli, A.~Bianconi, {Scale-free structural organization of oxygen interstitials in La$_2$CuO$_{4+y}$}.
\newblock {\it Nature\/} {\bf 466}, 841--844 (2010).

\bibitem{Bozin.2000}
E.~S. Bo{\v z}in, G.~H. Kwei, H.~Takagi, S.~J.~L. Billinge, {Neutron diffraction evidence of microscopic charge inhomogeneities in the CuO$_2$ plane of superconducting La$_{2-x}$Sr$_x$CuO$_{4}$} ($0 \leq x \leq 0.30$).
\newblock {\it Phys. Rev. Lett.\/} {\bf 84}, 5856 (2000).

\bibitem{Eisaki.2004}
H.~Eisaki, N.~Kaneko, D.~L. Feng, A.~Damascelli, P.~K. Mang, K.~M. Shen, Z.-X. Shen, M.~Greven, {Effect of chemical inhomogeneity in bismuth-based copper oxide superconductors}.
\newblock {\it Phys. Rev. B\/} {\bf 69}, 064512 (2004).

\bibitem{Castellan.2006}
J.~P. Castellan, B.~D. Gaulin, H.~A. Dabkowska, A.~Nabialek, G.~Gu, X.~Liu, Z.~Islam, {Two- and three-dimensional incommensurate modulation in optimally-doped Bi$_{2}$Sr$_{2}$CaCu$_{2}$O$_{8+{\delta}}$}.
\newblock {\it Phys. Rev. B\/} {\bf 73}, 174505 (2006).

\bibitem{Izquierdo.2006}
M.~Izquierdo, S.~Megtert, J.~P. Albouy, J.~Avila, M.~A. Valbuena, G.~Gu, J.~S. Abell, G.~Yang, M.~C. Asensio, R.~Com\'{e}s, {X-ray diffuse scattering experiments from bismuth-based high-$T_c$ superconductors}.
\newblock {\it Phys. Rev. B\/} {\bf 74}, 054512 (2006).

\bibitem{Poccia.2020}
N.~Poccia, S.~Y.~F. Zhao, H.~Yoo, X.~Huang, H.~Yan, Y.~S. Chu, R.~Zhong, G.~Gu, C.~Mazzoli, K.~Watanabe, T.~Taniguchi, G.~Campi, V.~M. Vinokur, P.~Kim, {Spatially correlated incommensurate lattice modulations in an atomically thin high-temperature Bi$_{2.1}$Sr$_{1.9}$CaCu$_{2.0}$O$_{8+y}$ superconductor}.
\newblock {\it Phys. Rev. Mater.\/} {\bf 4}, 114007 (2020).

\bibitem{Jakubowicz.2001}
N.~Jakubowicz, D.~Grebille, M.~Hervieu, H.~Leligny, {Simple and double modulations in Bi$_{2-x}$Pb$_{x}$Sr$_{2}$CaCu$_{2}$O$_{8+{\delta}}$}.
\newblock {\it Phys. Rev. B\/} {\bf 63}, 214511 (2001).

\bibitem{Beyers.1989}
R.~Beyers, B.~T. Ahn, G.~Gorman, V.~Y. Lee, S.~S.~P. Parkin, M.~L. Ramirez, K.~P. Roche, J.~E. Vazquez, T.~M. G{\" u}r, R.~A. Huggins, {Oxygen ordering, phase separation and the 60-K and 90-K plateaus in YBa$_2$Cu$_3$O$_x$}.
\newblock {\it Nature\/} {\bf 340}, 619--621 (1989).

\bibitem{Strempfer.2004}
J.~Strempfer, I.~Zegkinoglou, U.~R{\"u}tt, M.~v.~Zimmermann, C.~Bernhard, C.~T. Lin, T.~Wolf, B.~Keimer, {Oxygen superstructures throughout the phase diagram of $(\mathrm{Y},\mathrm{C}\mathrm{a})\mathrm{B}{\mathrm{a}}_{2}\mathrm{C}{\mathrm{u}}_{3}{\mathrm{O}}_{6+\mathrm{x}}$}.
\newblock {\it Phys. Rev. Lett.\/} {\bf 93}, 157007 (2004).

\bibitem{Axe.1994}
J.~D. Axe, M.~K. Crawford, {Structural instabilities in lanthanum cuprate superconductors}.
\newblock {\it J. Low Temp. Phys.\/} {\bf 95}, 271--284 (1994).

\bibitem{Bozin.1998}
E.~S. Bo{\v z}in, S.~J.~L. Billinge, G.~H. Kwei, {Re-examination of the second-order structural transition in La$_{2-x}$A$_x$CuO$_{4}$ (A = Ba, Sr)}.
\newblock {\it Physica B\/} {\bf 241--243}, 795 (1998).

\bibitem{Wakimoto.2006}
S.~Wakimoto, H.~Kimura, M.~Fujita, K.~Yamada, Y.~Noda, G.~Shirane, G.~Gu, H.~Kim, R.~J. Birgeneau, {Incommensurate lattice distortion in the high temperature tetragonal phase of La$_{2-x}$(Ba,Sr)$_x$CuO$_{4}$ }.
\newblock {\it J. Phys. Soc. Jpn.\/} {\bf 75}, 074714 (2006).

\bibitem{Yu.2019}
G.~Yu, D.-D. Xia, D.~Pelc, R.-H. He, N.-H. Kaneko, T.~Sasagawa, Y.~Li, X.~Zhao, N.~Bari{\v s}i{\' c}, A.~Shekhter, M.~Greven, {Universal precursor of superconductivity in the cuprates}.
\newblock {\it Phys. Rev. B\/} {\bf 99}, 214502 (2019).

\bibitem{Pelc.2019}
D.~Pelc, Z.~W. Anderson, B.~Yu, C.~Leighton, M.~Greven, {Universal superconducting precursor in three classes of unconventional superconductors}.
\newblock {\it Nat. Commun.\/} {\bf 10}, 2729 (2019).

\bibitem{Couture.2010}
G.~Chabot-Couture, {Synchrotron X-ray scattering studies of anomalous oxygen order in superconducting mercury barium copper oxide and of charge-transfer excitations in related undoped lamellar copper oxides}, Ph.d. thesis, Stanford University (2010).

\bibitem{Izquierdo.2011}
M.~Izquierdo, S.~Megtert, D.~Colson, V.~Honkim{\"a}ki, A.~Forget, H.~Raffy, R.~Com\'{e}s, {One dimensional ordering of doping oxygen in HgBa$_2$CuO$_{4+\delta}$ superconductors evidenced by x-ray diffuse scattering}.
\newblock {\it J. Phys. Chem. Solids\/} {\bf 72}, 545--548 (2011).

\bibitem{Welberry.2016}
T.~R. Welberry, D.~J. Goossens, {Interpretation of diffuse scattering in the high-$T_c$ superconductor HgBa$_2$CuO$_{4+\delta}$}.
\newblock {\it IUCrJ\/} {\bf 3}, 309 (2016).

\bibitem{Barisic.2013}
N.~Bari{\v s}i{\' c}, S.~Badoux, M.~K. Chan, C.~Dorow, W.~Tabis, B.~Vignolle, G.~Yu, J.~Beard, X.~Zhao, C.~Proust, M.~Greven, {Universal quantum oscillations in the underdoped cuprate superconductors}.
\newblock {\it Nat. Phys.\/} {\bf 9}, 761--764 (2013).

\bibitem{Barisic.2013a}
N.~Bari{\v s}i{\' c}, Y.~Chan, M K~Li, G.~Yu, X.~Zhao, M.~Dressel, A.~Smontara, M.~Greven, {Universal sheet resistance and revised phase diagram of the cuprate high-temperature superconductors}.
\newblock {\it Proc. Natl. Acad. Sci. U.S.A.\/} {\bf 110}, 12235--12240 (2013).

\bibitem{Gumbsch.2001}
P.~Gumbsch, S.~Taeri-Baghbadrani, D.~Brunner, W.~Sigle, M.~R{\"u}hle, {Plasticity and inverse brittle-to-ductile transition in strontium titanate}.
\newblock {\it Phys. Rev. Lett.\/} {\bf 87}, 085505 (2001).

\bibitem{Collignon.2019}
C.~Collignon, X.~Lin, C.~W. Rischau, B.~Fauqu{\'e}, K.~Behnia, {Metallicity and superconductivity in doped strontium titanate}.
\newblock {\it Annu. Rev. Condens. Matter Phys.\/} {\bf 10}, 25--44 (2019).

\bibitem{Gastiasoro.2020}
M.~N. Gastiasoro, J.~Ruhman, R.~M. Fernandes, {Superconductivity in dilute SrTiO$_3$ : A review}.
\newblock {\it Ann. Phys.\/} {\bf 417}, 168107 (2020).

\bibitem{Hameed.2020}
S.~Hameed, D.~Pelc, Z.~W. Anderson, R.~J. Spieker, M.~Lukas, Y.~Liu, M.~J. Krogstad, R.~Osborn, C.~Leighton, M.~Greven, {Enhanced superconductivity in plastically deformed strontium titanate}.
\newblock {\it Nat. Mater.\/} {\bf 21}, 54--61 (2022).

\bibitem{Khayr.2024}
I.~Khayr, S.~Hameed, J.~Budi{\'c}, X.~He, R.~S. Spieker, A.~Najev, Z.~Zhao, L.~Yue, M.~J. Krogstad, F.~Ye, Y.~Liu, R.~Osborn, S.~Rosenkranz, Y.~Li, D.~Pelc, M.~Greven, {Structural properties of plastically deformed SrTiO$_3$ and KTaO$_3$}.
\newblock {\it arXiv:2405.13249\/}  (2024).

\bibitem{Gay.1951}
P.~Gay, R.~W.~K. Honeycombe, {X-ray asterisms from deformed crystals}.
\newblock {\it Proc. Phys. Soc. A\/} {\bf 64}, 844--845 (1951).

\bibitem{Wang.2024}
X.~Wang, A.~Kundu, B.~Xu, S.~Hameed, N.~Rothem, S.~Rabkin, L.~Rogi{\'c}, L.~Thompson, A.~McLeod, M.~Greven, D.~Pelc, I.~Sochnikov, B.~Kalisky, A.~Klein, {Multiferroicity in plastically deformed SrTiO$_3$}.
\newblock {\it Nat. Commun.\/} {\bf 15}, 7442 (2024).

\bibitem{Fang.2024}
X.~Fang, J.~Zhang, A.~Frisch, O.~Preuß, C.~Okafor, M.~Setvin, W.~Lu, {Room‐temperature bulk plasticity and tunable dislocation densities in KTaO$_3$}.
\newblock {\it J. Am. Ceram. Soc.\/} {\bf 107}, 7054--7061 (2024).

\bibitem{Sears.2023a}
J.~Sears, Y.~Shen, M.~J. Krogstad, H.~Miao, J.~Yan, S.~Kim, W.~He, E.~S. Bo{\v z}in, I.~K. Robinson, R.~Osborn, S.~Rosenkranz, Y.-J. Kim, M.~P.~M. Dean, {Stacking disorder in $\alpha$-RuCl$_3$ investigated via x-ray three-dimensional difference pair distribution function analysis}.
\newblock {\it Phys. Rev. B\/} {\bf 108}, 144419 (2023).

\bibitem{Zhang.2024}
H.~Zhang, M.~A. McGuire, A.~F. May, H.-Y. Chao, Q.~Zheng, M.~Chi, B.~C. Sales, D.~G. Mandrus, S.~E. Nagler, H.~Miao, F.~Ye, J.~Yan, {Stacking disorder and thermal transport properties of $\ensuremath{\alpha}\text{\ensuremath{-}}{\mathrm{RuCl}}_{3}$}.
\newblock {\it Phys. Rev. Mater.\/} {\bf 8}, 014402 (2024).

\bibitem{Banerjee.2016}
A.~Banerjee, C.~A. Bridges, J.-Q. Yan, A.~A. Aczel, L.~Li, M.~B. Stone, G.~E. Granroth, M.~D. Lumsden, Y.~Yiu, J.~Knolle, S.~Bhattacharjee, D.~L. Kovrizhin, R.~Moessner, D.~A. Tennant, D.~G. Mandrus, S.~E. Nagler, {Proximate Kitaev quantum spin liquid behaviour in a honeycomb magnet}.
\newblock {\it Nat. Mater.\/} {\bf 15}, 733--740 (2016).

\bibitem{Banerjee.2017}
A.~Banerjee, J.-Q. Yan, J.~Knolle, C.~A. Bridges, M.~B. Stone, M.~D. Lumsden, D.~G. Mandrus, D.~A. Tennant, R.~Moessner, S.~E. Nagler, {Neutron scattering in the proximate quantum spin liquid RuCl$_3$}.
\newblock {\it Science\/} {\bf 356}, 1055--1059 (2017).

\bibitem{Takagi.2019}
H.~Takagi, T.~Takayama, G.~Jackeli, G.~Khaliullin, S.~E. Nagler, {Concept and realization of Kitaev quantum spin liquids}.
\newblock {\it Nat. Rev. Phys.\/} {\bf 1}, 264--280 (2019).

\bibitem{Cao.2016}
H.~B. Cao, A.~Banerjee, J.-Q. Yan, C.~A. Bridges, M.~D. Lumsden, D.~G. Mandrus, D.~A. Tennant, B.~C. Chakoumakos, S.~E. Nagler, {Low-temperature crystal and magnetic structure of $\alpha$-RuCl$_3$}.
\newblock {\it Phys. Rev. B\/} {\bf 93}, 134423 (2016).

\bibitem{Liu.2022}
Y.~Liu, H.~Cao, S.~Rosenkranz, M.~Frost, T.~Huegle, J.~Y.~Y. Lin, P.~Torres, A.~Stoica, B.~C. Chakoumakos, {\textit{PIONEER}, a high-resolution single-crystal polarized neutron diffractometer}.
\newblock {\it Rev. Sci, Instr.\/} {\bf 93}, 073901 (2022).

\bibitem{Koch.2021}
R.~J. Koch, N.~Roth, Y.~Liu, O.~Ivashko, A.-C. Dippel, C.~Petrovic, B.~B. Iversen, M.~v. Zimmerman, E.~S. Bo{\v z}in, On single-crystal total scattering data reduction and correction protocols for analysis in direct space.
\newblock {\it Acta Cryst. A\/} {\bf 77}, 611--636 (2021).

\bibitem{Welberry.2008}
T.~R. Welberry, D.~J. Goosens, The interpretation and analysis of diffuse scattering using monte carlo simulation methods.
\newblock {\it Acta Cryst. A\/} {\bf 64}, 23--32 (2008).

\bibitem{Weber.2002}
T.~Weber, H.-B. B{\"u}rgi, Determination and refinement of disordered crystal structures using evolutionary algorithms in combination with monte carlo methods.
\newblock {\it Acta Cryst. A\/} {\bf 58}, 526--540 (2002).

\bibitem{Nield.1995}
V.~M. Nield, D.~A. Keen, R.~L. McGreevy, The interpretation of single-crystal diffuse scattering using reverse monte carlo modelling.
\newblock {\it Acta Cryst. A\/} {\bf 51}, 763--771 (1995).

\bibitem{Welberry.1998}
T.~R. Welberry, T.~Proffen, Analysis of diffuse scattering from single crystals via the reverse monte carlo technique. i. comparison with direct monte carlo.
\newblock {\it J. Appl. Cryst.\/} {\bf 31}, 309--317 (1998).

\bibitem{Simonov.2014}
A.~Simonov, T.~Weber, W.~Steurer, \textit{Yell}: a computer program for diffuse scattering analysis via three-dimensional delta pair distribution function refinement.
\newblock {\it J. Appl. Cryst.\/} {\bf 47}, 1146--1152 (2014).

\bibitem{Morgan.2021}
Z.~J. Morgan, H.~D. Zhou, B.~C. Chakoumakos, F.~Ye, \textit{rmc-discord}: reverse monte carlo refinement of diffuse scattering and correlated disorder from single crystals.
\newblock {\it J. Appl. Cryst.\/} {\bf 54}, 1867--1885 (2021).

\bibitem{Pan.2023}
T.~Pan, S.~Jin, M.~D. Miller, A.~Kyrillidis, G.~N. Phillips, {A deep learning solution for crystallographic structure determination}.
\newblock {\it IUCrJ\/} {\bf 10}, 487--496 (2023).

\bibitem{Hamilton.2023}
P.~K. Hamilton, M.~Moya, Jaime, A.~M. Hallas, E.~Morosan, R.~Barala, B.~A. Frandsen, Symmetry-mode analysis for local structure investigations using pair distribution function data.
\newblock {\it J. Appl. Cryst.\/} {\bf 56}, 1192--1199 (2023).

\bibitem{Schmidt.2022}
E.~M. Schmidt, J.~M. Bulled, A.~L. Goodwin, Efficient fitting of single-crystal diffuse scattering in interaction space: a mean-field approach.
\newblock {\it IUCrJ\/} {\bf 9}, 21--30 (2022).

\end{thebibliography}
\bibliographystyle{ScienceAdvances}

\end{document}